\documentclass[12pt]{article}
\usepackage{graphicx,subfigure}
\usepackage{epstopdf}
\usepackage{amsmath}
\usepackage{amsfonts}
\usepackage{amssymb}
\newcommand{\beq}{\begin{equation}}
\newcommand{\eeq}{\end{equation}}
\newcommand{\be}{\begin{equation}}
\newcommand{\ee}{\end{equation}}
\newcommand{\beqa}{\begin{eqnarray}}
\newcommand{\eeqa}{\end{eqnarray}}
\newcommand{\beqar}{\begin{eqnarray*}}
\newcommand{\eeqar}{\end{eqnarray*}}
\newcommand{\bea}{\begin{eqnarray}}
\newcommand{\eea}{\end{eqnarray}}

\newcommand{\al}{\alpha}

\newcommand{\ch}{\textrm{ch}}
\newcommand{\sh}{\textrm{sh}}

\newcommand{\eg}{{\it e.g.,}\ }
\newcommand{\ie}{{\it i.e.,}\ }

\numberwithin{equation}{section}
 \allowdisplaybreaks[1]

\begin{document}

\setlength{\unitlength}{1mm}

\begin{titlepage}

\begin{flushright}
DCPT-08/63
\end{flushright}

\vspace*{.5cm}

\begin{center}
\LARGE{\bf  Black Rings in Taub-NUT\\
and D0-D6 interactions}
\end{center}

\vskip1cm

\begin{center}
{Joan Camps$^{a}$, Roberto Emparan$^{a,b}$, Pau Figueras$^{c}$,
Stefano Giusto$^{d}$, Ashish
Saxena$^{d}$}\\ \vspace{1cm}
$^a$\small{\textit{Departament de F\'{\i}sica Fonamental, Universitat de
Barcelona,\\ Mart\'{\i} i Franqu\`{e}s 1, E-08028, Barcelona,
Spain}}\\\vspace{1ex}
$^b$\small{\textit{Instituci\'{o} Catalana de Recerca i Estudis 
Avan\c cats (ICREA)\\
Passeig Llu\'{\i}s Companys, 23, E-08010 Barcelona, Spain}}\\
\vspace{1ex}
$^c$\small{\textit{Centre for Particle Theory, Department of
Mathematical Sciences, University of Durham,\\South Road, Durham, DH1
3LE, UK}}\\\vspace{1ex}
$^d$\small{\textit{Institut de Physique Th\'eorique, CEA/Saclay\\ 91191
Gif sur Yvette, France}}

\vskip.8cm 
\texttt{jcamps@ffn.ub.es, emparan@ub.edu, pau.figueras@durham.ac.uk, \\
stefano.giusto@cea.fr, ashish.saxena@cea.fr}

\end{center}

\date{}

\vspace{.5cm} 

\begin{abstract} We analyze the dynamics of neutral black rings in
Taub-NUT spaces and their relation to systems of D0 and D6 branes in the
supergravity approximation. We employ several recent techniques, both
perturbative and exact, to construct solutions in which thermal
excitations of the D0-branes can be turned on or off, and the D6-brane
can have $B$-fluxes turned on or off in its worldvolume. By explicit
calculation of the interaction energy between the D0 and D6 branes,
we can study equilibrium configurations and their stability. We find
that although D0 and D6 branes (in the absence of $B$ fields, and at
zero temperature) repeal each other at non-zero separation, as they get
together they go over continuosly to an unstable bound state of an
extremal singular Kaluza-Klein black hole. We also find that, for
$B$-fields larger than a critical value, or sufficiently large thermal
excitation, the D0 and D6 branes form stable bound states. The bound
states with thermally excited D0 branes are black rings in Taub-NUT, and
we provide an analysis of their phase diagram. \end{abstract}

\end{titlepage}

\setcounter{footnote}{0}

\tableofcontents

\newpage

\section{Introduction and Summary}

Recently there has been great progress in advancing techniques to
construct and analyze solutions for higher-dimensional black holes
\cite{ER3, NO1, VN1}. These black holes allow for non-spherical
topologies as well as extended horizons, and can often be related to
self-gravitating D-brane configurations. The progress has mainly come
from two different lines: (i) five-dimensional vacuum
solution-generating techniques have yielded many qualitatively new
solutions describing black rings and black holes \cite{ER4, ER1, ER2,
PS, saturn, diring, GS1, EK1, GS2, EMR1}; (ii) approximate methods have
allowed to construct and analyze thin black rings in a larger variety of
backgrounds and dimensions \cite{EHNOR1, CER1}.

In this paper we bring to bear, and at some points refine and extend,
these techniques to analyze black rings in backgrounds of Taub-NUT type.
Such solutions describe, when embedded in M-theory and then reduced to
IIA theory, D0-branes in the presence of D6-branes. The self-gravitating
D6-brane is essentially a Kaluza-Klein (KK) monopole plus six additional
space dimensions of M-theory, while the D0 brane uplifts to a momentum
wave of gravitons along the eleventh direction. When the D0 is excited
thermally, this graviton wave develops a horizon and becomes a boosted
black string --- conversely, when the boost becomes light-like and the
horizon shrinks to zero, we recover the gravitational description of a
D0 brane in its supersymmetric ground state. In the Taub-NUT background,
where the eleven-dimensional direction is contractible, the black string
is actually a black ring. So, quite generally, arrangements of D0 and D6
branes are described in supergravity as black rings in Taub-NUT. As
four-dimensional solutions, they describe a (singular) magnetic monopole
and an electrically charged black hole. The KK magnetic and electric
charges, $P$ and $Q$, are related to the numbers $N_6$ and $N_0$ of D6
and D0 branes through the length of the compact Kaluza-Klein circles
near asymptotic infinity, $2\pi L$,
\beq\label{PQN}
P=\frac{LN_6}{4}\,,\qquad Q=\frac{2G_4 N_0}{L}\,.
\eeq
For the most part we will consider for simplicity a single D6 brane,
$N_6=1$.
Configurations of this sort have been constructed in the supersymmetric
case in \cite{EEMR1, BKW1,GSY1}. Our aim is to study the much more complex
non-supersymmetric D0-D6 systems.

The D0-D6 system presents a number of peculiar features. The
long-distance D0-D6 interaction, mediated by NSNS gravi-dilaton and RR
gauge-field closed-string exchange, is repulsive. This admits a simple
interpretation in M-theory, where the effect is simply the centrifugal
force created by the rotation of a ring with light-like local boost. It
must be noted, though, that despite this long-distance repulsion, D0
branes can bind to the worldvolume of D6 branes to form
non-supersymmetric metastable bound states \cite{WT1} which, at strong
coupling, can be precisely matched to Kaluza-Klein black holes with non-zero
Bekenstein-Hawking entropy \cite{DR1, FL1,EH1, EM}. 

It is known that the physics of D0-D6 interactions becomes
richer when $B$-form fluxes, introduced as moduli, are turned on in the
worldvolume of the D6 \cite{Chen,Mihailescu,EW1}. When the fluxes are large enough (more precisely,
when a codimension-1 wall is crossed in the moduli space of $B$-fields)
it is possible to have supersymmetric bound states of D0 and D6 \cite{Chen,EW1}, where the D0 is at a finite distance from the D6. 
But we can envisage another way of achieving equilibrium between a set
of D0 branes and a D6 brane (without $B$-fields). If we add some energy
of excitation to the D0 branes while keeping their charge fixed, we
enhance the gravitational attraction to the D6 brane, which may then
overcome their repulsion. When the excitations of the D0 branes (\ie
of the open strings stretched between a gas of D0s) have a
thermal distribution, then in the regime of validity of supergravity
they are described as D0-charged black holes. Thus, if the horizon area
of the D0-charged black hole is large enough, a non-supersymmetric bound
state may be possible. A main aim of this paper is to demonstrate these
two mechanisms using the novel gravity techniques mentioned above.

Finding equilibrium configurations is not the only information we can
obtain from our methods: we can also study their stability and their
interaction energy. In constrast to the techniques based on solving
Killing spinor equations, we can construct configurations in which the
separation $R$ between D0 and D6 branes does not correspond to
equilibrium (so supersymmetry is broken). We compute the interaction
energy as the difference between the total ADM energy of the system, as
measured at infinity, and the masses of the D0 and D6 branes when they
are isolated from each other,
\beq\label{eintdef}
E_\mathrm{int}(R)=M_\mathrm{tot}(R)-M_\mathrm{D0}-M_\mathrm{D6}\,.
\eeq
The D0-D6 separation in equilibrium states corresponds to extrema of
this energy for fixed charges and horizon area, essentially as a
consequence of the first law. \textit{Stable} configurations should
correspond to minima. When the D0s are not excited and so have zero
entropy, $M_\mathrm{D0}$ is simply determined by its charge (\textit{i.e.}, net
number of D0 branes). When the D0s are thermally excited, we take
$M_\mathrm{D0}$ to be the mass of a D0-charged black hole with fixed
values of the charge and area (entropy). Thus $E_\mathrm{int}$ measures
the interaction energy as the thermally-excited D0 branes are moved
adiabatically towards the D6 brane\footnote{Alternatively, we could fix
the energy of the D0s, in which case the configuration of stable
equilibrium would be that which maximizes the entropy. The results are
in both cases essentially the same, and for definitiness we choose the
fix the area in order to determine $M_\mathrm{D0}$.}. For the
perturbative solutions a convenient alternative way to determine the
stability is to analyze the external force needed to balance the
configurations away from equilibrium--- a potential associated to this
force can also be constructed, which is closely related to
$E_\mathrm{int}$. 

\begin{figure}[h!]
\label{fig:D0D6noB}
\centering
\includegraphics[scale=0.9]{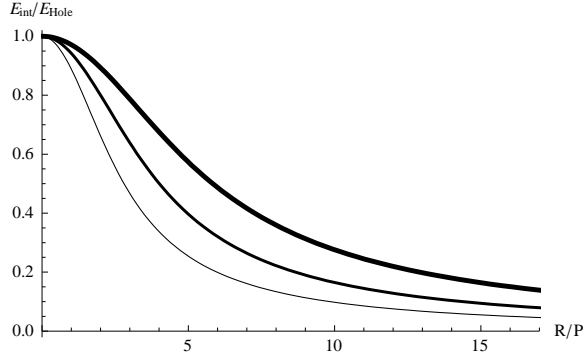}
\hspace{0.5cm}
\caption{\small{D0-D6 interaction energy $E_\text{int}$ (with no $B$-field and
at zero temperature) as a function of separation $R$, for $Q/P=.1$
(thin), $1.$ (thick) and $10$ (thicker). The interaction energy is
normalized relative to the one of the (singular) extremal black hole
with the same electric and magnetic charges, and angular momentum
$G_4J=PQ$. Here $E_\text{int}$ is computed using the exact solutions of
sec.~\ref{sec:exactextr}.}
}
\end{figure}

\begin{figure}[h!]
\centering
\includegraphics[scale=0.9]{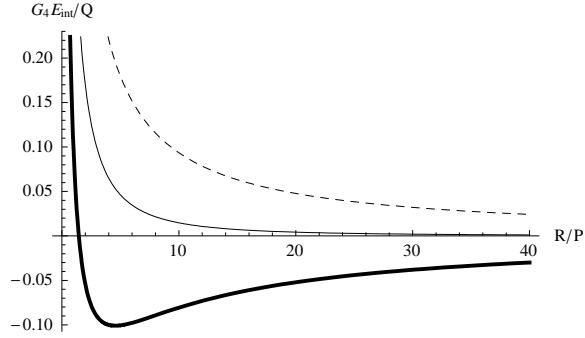}
\hspace{0.5cm}
\caption{\small{D0-D6 interaction potential with flux $B=bL/2$ versus $R$, for
$b=1/9$ (dashed), $b_c=1/\sqrt{3}$ (thin solid) and
$.8$ (thick solid). The
interaction energy is normalized with the electric charge so it
corresponds to a fixed net number of D0 branes. The potential is
obtained from the perturbative extremal solutions of
sec.~\ref{sec:pertd0d6}.}}
\label{fig:D0D6B}
\end{figure}

\begin{figure}[h!]
\centering
\includegraphics[scale=0.9]{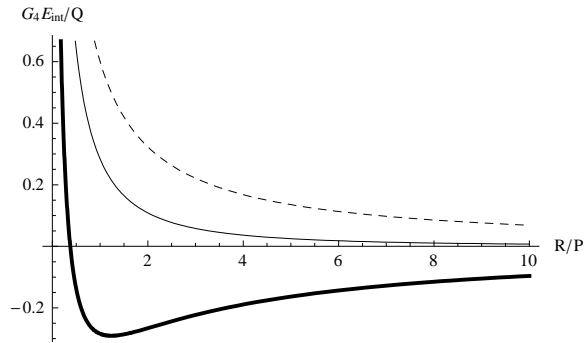}
\hspace{0.5cm}
\caption{\small{Thermal D0-D6 interaction potential versus $R$, for fixed
values of the entropy and charge,
$G_4 S/(4\pi Q^2)=.1$ (dashed), critical $1/\sqrt{2}$ (thin solid) and
$3.$ (thick solid). The
interaction energy is normalized with the electric charge. The potential
is obtained from the perturbative non-extremal solutions of
sec.~\ref{sec:pertd0d6} with $b=0$.}}
\label{fig:D0D6therm}
\end{figure}

Our main results are: 

\begin{enumerate}

\item In the absence of $B$-fields, the closed-string interaction
between the D0 and D6 at any finite separation is repulsive. However,
the interaction energy goes continuously to a finite maximum as the
distance between the D0 and the D6 decreases to zero. We construct a
family of exact solutions that in this limit describe the formation of
an unstable D0-D6 bound state
corresponding to an extremal (singular) Kaluza-Klein black hole with
angular momentum $J=PQ/G_4$ and mass
\beq\label{bhmass}
M_\mathrm{bh}=E_\mathrm{int}+M_\mathrm{D0}+M_\mathrm{D6}\,.
\eeq 
See figure \ref{fig:D0D6noB}. To obtain this result it is crucial to
work with the exact solutions: perturbative calculations break down as
the D0 and D6 get together, since they give
$E_\mathrm{int}\to\infty$.

\item When $B$-fields are turned on in the D6 worldvolume, the interaction
is again repulsive below a critical value $B_c$, but for
$B\geq B_c=L/2\sqrt{3}$ a stable minimum develops. The equilibrium solutions we
find, using an approximate construction of thin black rings, reproduce
precisely previous results based on rather different,
supersymmetry-based, techniques. Our methods also provide the
off-shell interaction energy and thus a simple way to check the
stability of these configurations. See figure \ref{fig:D0D6B}.

\item Thermally excited D0-branes, even in the absence of $B$-fluxes,
can achieve equilibrium configurations in the presence of D6 branes if
the entropy of the D0-brane thermal gas is larger than a critical value
\beq
S>S_{c}=8\sqrt{2}\pi N_0^2\,\frac{G_4}{L^2}\,,
\eeq
or equivalently, if the mass of the excited D0s is
\beq
M>M_c=\frac{3}{\sqrt{2}}\frac{N_0}{L}\,.
\eeq 
These are proper black rings in Taub-NUT space, in the sense that they
have regular horizons of finite area. Again, we construct both exact and
perturbative solutions for such systems. For the perturbative solutions
we obtain $E_\mathrm{int}$, which allows to argue the stability of the
bound states under changes in the distance between the D0 and D6, see
figure \ref{fig:D0D6therm}. The exact solutions that we construct are
only a subfamily of the most general class of exact solutions for black
rings in Taub-NUT, since we cannot vary independently the $S^1$ and
$S^2$ angular momenta of the black ring. 

\end{enumerate}

Point 1 above deserves further comment. As we have mentioned,
ref.~\cite{WT1} described how D0 branes can bind to the worldvolume of
D6 branes and form quadratically (meta-)stable bound states. The
construction in \cite{WT1} did not include any angular momentum.
However, one expects that angular momentum can be added in the form of
fermionic excitations of the 0-6 open strings. When these fill up to the
Fermi level, the configuration will have angular momentum $J=N_0N_6/2$,
and vanishing macroscopic entropy. This is precisely like in the
extremal Kaluza-Klein black hole with $J=PQ/G_4$. Our result
\eqref{bhmass} amounts to an \textit{exact} computation of the mass of
this bound state by taking into account the energy of closed-string
interaction stored in the bound state as $N_0$ D0 branes are moved
towards $N_6$ D6 branes.\footnote{Observe that given our definition
\eqref{eintdef} and that $M_\mathrm{bh}$ is also an ADM mass, what makes
this result non-trivial is that the black hole solution can be reached
continuously from the solutions for separate D0 and D6 branes.}
Moreover, the fact that the exact interaction energy reaches a maximum
that accounts precisely for the mass of the black hole strongly suggests
that, for this limiting value of the angular momentum, the state of the
D0 branes in the worldvolume of the D6 brane with $J=N_0N_6/2$ is not a
metastable minimum as the one with $J=0$ in \cite{WT1}\footnote{Whose local
potential is created by open string interactions.} but is actually an
unstable maximum. It would be interesting to derive this result from an
analysis of the D6 worldvolume gauge theory.

Regarding point 3, we remark that the entropy and mass of the thermal
D0s will not only be bounded below for given $N_0$, but they will also
be bounded {\it above}. This is, if the D0 branes become too massive,
the repulsive effect between D0 and D6 charges will be overwhelmed by the
attraction between their masses. In terms of black rings, this
corresponds to the fact that the mass and area of a black ring with a
given spin along the $S^1$ are bounded above, the upper values
corresponding to the solutions where the thin and fat branches of black
rings meet. This regime is away from our perturbative techniques, and we
cannot obtain the precise dependence of these upper bounds on $N_0$. For
very small D0 charge, however, the values can be approximately obtained
from the asymptotically flat case,
\beq\label{upperbounds}
S\leq \frac{4\pi}{3\sqrt{3}}N_0\,,\qquad
M\leq \left(\frac{N_0}{\sqrt{2G_4 L}}\right)^{2/3}\qquad
(G_4 M\ll L)\,.
\eeq

\bigskip

The paper is structured as follows: section \ref{sec:pertapp} develops
the general technique for approximate perturbative solutions of thin
black rings. Here we follow and expand on \cite{EHNOR1,CER1}. In
section~\ref{sec:pertd0d6} this technique is applied to put D0 branes in
the background of a D6 brane, possibly with $B$ fluxes on the D6 and
finite temperature on the D0s. This yields the equilibrium
configurations discussed in points 2 and 3 above and the results for the
interaction energy plotted in figs.~\ref{fig:D0D6B} and
\ref{fig:D0D6therm}. Section~\ref{constructionsec} describes the exact
solution-generating method (following \cite{GS1, GS2}) and then proceeds
to generate the basic solution for a black ring in Taub-NUT. In section
\ref{sec:exactextr} we particularize to extremal D0 branes, and compute
the interaction energy presented in fig.~\ref{fig:D0D6noB}. We also
exhibit the limit, discussed in point 1 above, to an extremal KK black
hole as the separation between the D0 and D6 vanishes. Section
\ref{sec:exactrings} studies a family of exact black rings in Taub-NUT,
and discusses their phase diagram. We conclude in section
\ref{sec:outlook}.

\bigskip
\noindent
\textit{Note added:} H.~Yavartanoo has informed us of the paper \cite{yavar}, which deals with
related issues.

\section{Perturbative approach to thin rings: General method}
\label{sec:pertapp}

The first approach is based on the method developed in \cite{EHNOR1} for a
systematic perturbative construction of thin black rings in a background
that possesses a $U(1)$ isometry. The ring lies along an orbit of the
isometry, and at the zero-th order level of approximation that we work
on in this paper, its backreaction on the geometry will be neglected.

The method requires the existence of two widely separated scales, one of
which is the ring's $S^2$-thickness, $r_0$, and the other one is a large
scale $R\gg r_0$ that is typically either a measure of the ring's radius
or a characteristic length scale of the background --- whichever of the
two is smaller. The method allows to determine readily the range of
validity of the approximation. At scales much larger than $r_0$, we can
obtain the linearized field created by the ring by substituting it with
a distributional source of energy-momentum $T_{\mu\nu}$.

Ref.~\cite{BC1} showed that when a brane with distributional energy
momentum $T^{\mu\nu}$
and worldvolume spanning a submanifold of extrinsic curvature
tensor ${K_{\mu\nu}}^\sigma$, is subject to an external force
density $\mathcal{F}^\sigma$ along a direction transverse to its
worldvolume,
then it must satisfy the equations of motion
\beq\label{cartereom}
\mathcal{F}^\sigma=T^{\mu\nu}{K_{\mu\nu}}^\sigma \,.
\eeq
In the absence of external forces, $\mathcal{F}^\sigma=0$, this equation
imposes a constraint on the sources one can place in a given curved
submanifold. In the case of a ring, the circle where the ring lies
typically has non-zero extrinsinc curvature, so the equation
\beq\label{cartereom2}
T^{\mu\nu}{K_{\mu\nu}}^\sigma=0
\eeq 
determines the value of the rotation (locally a boost) for which the
centripetal ring tension and the centrifugal repulsion balance each
other --- recall that the gravitational self-interaction of the ring is
neglected in this approximation.

By analyzing how this force changes as we change the ring radius, we
obtain information about the radial stability of a ring. If to increase the ring radius we need to apply an outward-pushing force, then the ring
will be radially stable. If instead we have to push inward to keep the ring in a sligthly larger radius, the equilibrium will be unstable \cite{EEV1}.\footnote{Bear
in mind that the force depends on the radius through the geometry,
${K_{\mu\nu}}^\sigma$, but also possibly through the ring's parameters in
$T_{\mu\nu}$, since typically we will want to keep a physical parameter
(such as mass, charge, or
area) fixed as the radius is varied.} The same
information can be put in a perhaps more convenient way if we first integrate
the force \eqref{cartereom} along the ring's radial direction, to obtain
a potential for the ring in this background. The two situations
described above then correspond to minima and maxima of this potential.

\subsection{Local analysis}
\label{subsec:local}

We take the
isometry of the background along which the ring lies as being parametrized by
a coordinate $z$. The location of the
ring can be conveniently specified as the zero of some coordinate $\rho$
measuring radial distance transverse to the circle. Then,
close to $\rho=0$ we can always write the background geometry,
to lowest order in
$\rho/R$, as flat space in the form
\beq\label{zerothbckg}
ds^2=-d\tau^2+dz^2+d\rho^2+\rho^2(d\vartheta^2+\sin^2\vartheta
d\phi^2)+O(\rho/R)\,.
\eeq
In this background, we place a distributional source of energy-momentum
that reproduces the asymptotic field created by a black {\it string} in
a flat space background. For a boosted black string, this is
\begin{eqnarray}\label{tensorcomponents}
&&T_{\tau\tau}=\frac{r_0}{16\pi G_5}\left(\cosh^2\al+1\right)\;\delta^{(3)}(\rho)\,,\nonumber\\
&&T_{\tau  z}=\frac{r_0}{16\pi G_5}\;\cosh\al\sinh\al\;\delta^{(3)}(\rho)\,,\\
&&T_{z z}=\frac{r_0}{16\pi G_5}\left(\sinh^2\al-1\right)\;\delta^{(3)}(\rho)\,.\nonumber
\end{eqnarray}
One case of particular interest is the extremal limit in which the
boost becomes light-like, $\alpha\to\infty$ with $r_0\cosh^2\alpha=p$ finite, 
\beq\label{extremaltensor}
T_{\tau\tau}=T_{\tau z}=T_{z z}=\frac{p}{16\pi G_5}\;\delta^{(3)}(\rho)\,.
\eeq

Since $T_{\tau  z}$ gives the momentum carried by the string along
$z$, and we assume that $z$ is a periodic coordinate, $z\sim z+\Delta
z$,  in a quantum
theory the parameters will be quantized,
\beq\label{quantN0}
\frac{r_0}{8\pi G_5}\cosh\alpha\sinh\alpha (\Delta z)^2=N_0\,,
\eeq
with integer $N_0$. In the context of this paper, in which the direction
along the string is
dimensionally reduced to obtain IIA solutions, the integer $N_0$
typically corresponds to the net number of D0 branes, and possibly a
contribution to the (quantized) four-dimensional angular momentum.

As mentioned above, the ring must satisfy the equations of motion
\eqref{cartereom}.
Since these are local equations at the position of the
ring, we can analyze them most easily and most generally by considering
the spacetime geometry close to the ring. The extrinsic curvature of the
ring's circle is a $O(1/R)$ effect, and thus to account for it we must go beyond
the zero-th order background \eqref{zerothbckg} and include corrections
to first order in $\rho/R$. A wide class of backgrounds are covered by
considering corrections of the form
\beqa\label{genback}
ds^2&=&-\left(1+C_{\tau\tau}\frac{2\rho\cos\vartheta}{R}\right)d\tau^2
+\left(1+C_{zz}\frac{2\rho\cos\vartheta}{R}\right)dz^2
+2C_{\tau z}\frac{2\rho\cos\vartheta}{R}d\tau dz\nonumber\\
&&+2 C_{\tau\phi}\frac{2\rho\sin\vartheta}{R} \rho\sin\vartheta d\phi d\tau
+2 C_{z\phi}\frac{2\rho\sin\vartheta}{R} \rho\sin\vartheta d\phi dz\nonumber\\
&&+\left(1+C_{\rho\rho}\frac{2\rho\cos\vartheta}{R}\right)
\left(d\rho^2+\rho^2d\vartheta^2+\rho^2\sin^2\vartheta d\phi^2\right)
+O(\rho^2/R^2)
\,,
\eeqa
where $C_{\mu\nu}$ are constants that are determined by the specific embedding
of the circle in the background. The Riemann tensor of this geometry
actually vanishes up to terms $O(R^{-2})$. Relative to the analysis in
\cite{EHNOR1}, we have added new crossed terms $C_{\tau\phi}$ and $C_{z\phi}$
which can appear at the same order. Eq.~\eqref{genback} describes the most
generic class of backgrounds deformed by $S^2$-dipole perturbations that
preserve the isometries generated by $\partial_\tau$, $\partial_z$ and
$\partial_\phi$. The value of $C_{\rho\rho}$ can be adjusted at will by
an appropriate gauge choice. 

In this background, we place a black string of thickness $r_0$, which
will modify the geometry \eqref{genback} at distances $\rho\sim r_0$.
Thus the approximations we use will be valid as long as
\beq\label{thinlimit}
r_0\ll \min\left(R,\frac{R}{|C_{\mu\nu}|}\right)\,.
\eeq

Eq.~\eqref{cartereom} can be readily evaluated in \eqref{genback}, since
in this case the
extrinsic curvature is simply
\be
K_{\mu\nu\sigma}=-\frac{1}{2}\partial_\sigma g_{\mu\nu}
\ee
where $\mu, \nu$ are parallel to the string ($\tau$ and $z$) and $\sigma$
is perpendicular. Since, to first order in $\rho/R$ the only coordinate
dependence is of dipole type we only have derivatives of the dipolar
contribution, $\rho\cos\vartheta$, which evaluated at the location of
the string $(\rho=0,\vartheta=0)$ are purely radial and along the plane
in which the ring is curved,
\be\label{force}
\mathcal{F}=\frac{1}{R}\left(C_{\tau\tau}T_{\tau\tau}+2C_{\tau z}T_{\tau z}-
C_{zz}T_{zz}\right)d\rho\,.
\ee
An equilibrium configuration is one for which
\beq\label{noforce}
C_{\tau\tau}T_{\tau\tau}+2C_{\tau z}T_{\tau z}=
C_{zz}T_{zz}\,.
\eeq
In the particular case in which the string is boosted to the speed of
light, \eqref{extremaltensor}, this reduces to
\beq\label{equilight}
C_{\tau\tau}+2C_{\tau z}=C_{zz}\,.
\eeq
It is easy to check that this is the same as the equation that
determines the null geodesics in \eqref{genback} on the plane
$\vartheta=0$ and at fixed radius $\rho\to 0$. In this case the results
are equivalent to more conventional massless probe calculations. But to
include non-extremal cases we must resort to the more general approach
described above.

\subsection{Physical magnitudes}
\label{sec:physmag}

We assume the existence of two commuting Killing vectors that correspond
to the canonically-normalized generators of time translation, $\zeta$,
and spatial $U(1)$ isometry at infinity, $\chi$. These are related by
linear combination to the Killing vectors $\partial_\tau$ and
$\partial_z$ in the region close to the ring,\footnote{Note that $\zeta$
cannot appear in the relation between $\chi$ and $\partial_z$ unless we
introduce closed timelike curves. Also, in principle
other isometries, such as $\partial_\phi$, may mix: the discussion below
can be easily modified to accommodate this.}
\beqa
\partial_\tau&=&a_0 \zeta +b_0\chi\,,\nonumber\\
\partial_z&=&b_1 \chi\,.
\eeqa
The coefficients $a_i$, $b_i$, reflect the possible
redshift between the vicinity of the ring and asymptotic infinity, as
well as possible rotations and twists between these two regions. 

The quantities conjugate to these Killing vectors are obtained through
integrals of the stress-energy tensor on a spacelike surface $\Sigma$
with normal $n^\mu$ and volume element $dV$.
The mass is conjugate to $\zeta$, so we define
\beq\label{genmass}
M=\int_{\Sigma} dV  T_{\mu\nu}n^\mu\zeta^\nu\,,
\eeq
and the momentum conjugate to the $U(1)$ isometry
\beq\label{genmomentum}
J=\int_{\Sigma} dV  T_{\mu\nu}n^\mu\chi^\nu\,
\eeq
(often this is an
angular momentum, but in some instances it is better regarded as linear
momentum or electric charge). We can readily obtain the surface gravity and horizon velocity associated
to $\zeta$ and $\chi$ from knowledge of the surface gravity and velocity
for $\partial_\tau$ and $\partial_z$. The Killing generator of the horizon is
\beq
\hat\xi=\partial_\tau+\tanh\alpha\partial_z=a_0\xi
\eeq
where 
\beq
\xi=\zeta+\frac{b_0+b_1\tanh\alpha }{a_0}\chi
\eeq
is the horizon generator in terms of the canonical asymptotic symmetry
generators $\zeta$ and $\chi$. Thus the horizon velocity relative to
infinity is
\beq
\Omega_H=\frac{b_0+b_1\tanh\alpha }{a_0}\,.
\eeq
The surface gravity associated to $\hat\xi$ is 
\beq
\hat\kappa=\frac{1}{2r_0\cosh\alpha}
\eeq
and so the surface gravity measured by asymptotic observers that follow
orbits of $\xi$ is
\beq
\kappa=\frac{1}{ 2a_0r_0\cosh\alpha}\,.
\eeq

Finally, the horizon area is computed as the area of the boosted black string,
\beq
\mathcal{A}_5=4\pi r_0^2 \Delta z \cosh\alpha\,.
\eeq

In all examples considered so far, these magnitudes are seen to satisfy
a first law
\beq
dM=\frac{\kappa}{8\pi G_5}d\mathcal{A}_5 +\Omega_H dJ
\eeq
for variations among stationary solutions
when, and only when, the equilibrium conditions are satisfied. We
believe this should be
generic, and provides a justification for the definitions
\eqref{genmass} and \eqref{genmomentum}.

\section{D0-D6 interaction: perturbative methods}
\label{sec:pertd0d6}

We now apply the methods of sec.~\ref{sec:pertapp} to study the
interaction between D0 and D6 branes in the supergravity
approximation. In the absence of other charges, fluxes, or excitations,
D0 and D6 branes repel each other. This should be reflected in the
non-existence of a supergravity solution that describes them in
equilibrium. More precisely, a solution in which the D0 and D6 remain
static at a finite distance from each other must contain external forces
holding them in place.

However, we expect two ways in which D0 and D6 branes may form bound
states at finite separation. The first one has been studied in some
detail in the past: a D0 and a D6 brane can form a supersymmetric
configuration if an appropriate $B$-field is turned on in the worldvolume
of the D6. When the $B$-field, which is a modulus, is above a critical
value $B_c$, a bound state between the D0 and D6 appears \cite{Chen,EW1}. A way
to understand this effect is by observing that the $B$-field on the D6
worldvolume induces, through the worldvolume Chern-Simons coupling,
Ramond-Ramond fields giving rise to D0, D2, D4 charges. The D2 branes
have an attractive effect on the D0, and if the $B$-field is large
enough this attraction may compensate the D6 repulsion. We will be able
to study the interaction energy and make explicit how, as
the $B$-field modulus is varied, the potential changes from having no
minimum when $B<B_c$, to developing one for $B>B_c$.\footnote{A threshold case with $B=B_c$ was studied in \cite{GMS1}.}

A second way in which D0 and D6 branes can be expected to overcome
their repulsion is
by turning on excitations that increase their gravitational attraction.
For simplicity we will only consider thermal excitations of the D0
brane, but in principle it is also possible (and not much more
difficult) to excite the D6 brane. Thus we consider a gas of D0 branes
with open strings stretching among them in a thermal ensemble. At
weak coupling and low energies, this is described by Super-Yang-Mills
quantum mechanics at finite temperature, and at higher energies in terms
of long excited strings with endpoints on the D0 branes.
At strong
coupling the description is in terms of a black hole with D0 charge,
which in M-theory terms is a black string boosted along the eleventh
direction. In the presence of a D6 brane, this becomes a black ring in
Taub-NUT.

As in the rest of the paper, we shall take the, more geometrical,
M-theory point of view on the system, and thus consider the D0-brane
uplifted to M-theory. In a probe approximation, the D0 brane is usually
studied as a massless particle moving in a geometry with the structure
of a Taub-NUT geometry. However, as discussed at the end of
sec.\ref{subsec:local}, in order to include also thermal excitations of
the D0 brane we must resort to the thin black ring approach
developed above. 

\subsection{Background D6 with $B$-flux}

The background in which we place the D0 brane is that of a D6-brane
wrapped on $T^6$ with a $B_{ab}$ field along its worldvolume
directions. For simplicity we shall consider the most symmetric
configuration where $B_{12}=B_{34}=B_{56}$, and with a single D6 brane.
The construction of the solution, uplifted to M-theory and reduced on
$T^6$ down to five dimensions, is detailed in
appendix~\ref{app:d6b}. For the purpose of studying the black ring
(\ie the M-uplifted D0 brane) in this background, we shall only need
the five-dimensional background metric
\beqa\label{d6b}
ds^2&=&-Z^{-2}\left[dt+\omega_0 \left(d\psi+(\cos\theta-
1)d\phi\right)\right]^2\nonumber\\
&&+\frac{Z}{H}\left(d\psi+(\cos\theta-
1)d\phi\right)^2+ZH(dr^2+r^2d\theta^2+r^2\sin^2\theta d\phi^2)
\eeqa
with $\psi\sim\psi+4\pi$, and where
\beq
H=h+\frac{1}{r}\,,\qquad Z=h_q+\frac{h_p^2}{H}\,,
\eeq
and
\beq
\omega_0 =\frac{3 h_p\;h_q}{2 H}+\frac{h_p^3}{H^2}\,.
\eeq
The moduli at infinity $h$, $h_p$, $h_q$ are given by 
the asymptotic Kaluza-Klein radius $L$ and the $B$-field, which we
express in terms of a dimensionless parameter $b$ as $B=bL/2$
\beq\label{hhphq}
h=\frac{2}{L}\frac{1-3b^2}{(1+b^2)^{3/2}}\,,\qquad
h_p=\frac{2b}{\sqrt{1+b^2}}\,,\qquad h_q=\frac{L}{2}\sqrt{1+b^2}\,.
\eeq

When the $B$-field vanishes, $b=0$, we recover the conventional
Kaluza-Klein monopole background,
with
\beq
H=\frac{2}{L}+\frac{1}{r}\,,\qquad Z=\frac{L}{2}\,,\qquad
\omega_0=0\qquad (b=0)\,.
\eeq
It is well known that this geometry is smooth at the core of the
KK monopole ($r=0$). The same is true
for generic values of $b$. The moduli induce
D0-D2-D4 charges, but these do not grow a horizon around the nut (which
would require charges not induced by the $B$-field), nor create a
singularity.

We now apply the methods of sec.~\ref{sec:pertapp} to this background.
First we need the form of the geometry near the location of the
ring, which we take to be the circle at 
\beq\label{ringcircle}
r=R\,,\qquad \theta=0\,
\eeq
extended along $\psi$.
Let us denote
\beq
H_R\equiv H(r=R)\,,\qquad Z_R\equiv Z(r=R)\,,\qquad \omega_R\equiv \omega_0(r=R)\,.
\eeq

The proper circumferential length of
the circle is
\beq\label{Delz}
\Delta z =\int d\psi \sqrt{g_{\psi\psi}}|_{r=R,\theta=0}
=2\pi L\frac{\sqrt{2R\left((1+b^2)^{3/2}L+2R\right)}}{\sqrt{1+b^2}L+2R}\,.
\eeq
As the distance between the ring and the nut grows, $R\to\infty$, this
becomes equal to the asymptotic KK circle length
\beq
\Delta z\to 2\pi L\,.
\eeq
Observe that $R$ is not the proper radial distance between the ring and
the nut but only a coordinate distance associated to the conventional
cooodinate $r$ in \eqref{d6b}. However, we will continue to use it as a
simple and convenient measure of the separation between the D0 and D6
brane.

In order to focus on the region around the circle \eqref{ringcircle}, we
change to adapted coordinates $(r,\theta)\to(\rho,\vartheta)$
\beq
r \sin\theta=\frac{\rho}{\sqrt{H_R Z_R}}\sin\vartheta,\qquad r
\cos\theta= R+\frac{\rho}{\sqrt{H_R Z_R}}\cos\vartheta
\eeq
such that the ring circle \eqref{ringcircle} now lies at $\rho=0$,
and then expand the metric in powers of $\rho/R$. To bring the metric
into the form \eqref{zerothbckg} to zero-th order in $\rho/R$, we have
to perform two further coordinates changes: first, change to corotating
coordinates, and then rescale time to canonical normalization and $\psi$
to proper length
along the string direction,
\beq\label{chKill}
t= \frac{\Delta z}{4\pi}\sqrt{Z_RH_R}\;\tau,\qquad
\psi=\frac{4\pi}{\Delta z}\left(z+ \sqrt{\frac{H_R}{Z_R^3}} \omega_{R}\;\tau\right).
\eeq

Now the metric to first order in $\rho/R$ takes the form \eqref{genback}, with
\beq\label{cmunu}
\begin{split}
C_{\tau\tau}=&\frac{L b^2}{(1+b^2)^{3/2}L+2R}\sqrt{\frac{2(1+b^2)R^3}{\left(\sqrt{1+b^2}L+2R\right)^3}}\\
C_{\tau z}=&\frac{2b^3 L R^2}{\left(\sqrt{1+b^2}L+2R\right)^2\left((1+b^2)^{3/2}L+2R\right)}\\
C_{zz}=&\frac{L\left((1+b^2)^{3/2}L+2(1-b^2)R\right)}{(1+b^2)^{3/2}L+2R}\sqrt{\frac{(1+b^2)R} {2\left(\sqrt{1+b^2}L+2R\right)^3}}\\
C_{\rho\rho}=&-L\sqrt{\frac{(1+b^2)R}{2\left(\sqrt{1+b^2}L+2R\right)^3}}\\
C_{z\phi}=&-\frac{L\sqrt{2R\left((1+b^2)^{3/2}L+2R\right)}}{4\left(\sqrt{1+b^2}L+2R\right)^2}\,,
\end{split}
\eeq
and $C_{\tau\phi}=0$.

The thin ring approximation is valid when \eqref{thinlimit} holds.
When $b=0$ this condition is always parametrically equivalent to simply
\beq\label{thinvalid}
r_0\ll R\,.
\eeq
Observe that when the ring is far from the nut, $R\gg L$, the
approximation is also valid for $r_0>L$: in this regime, in which the
black ring is very well approximated by a wrapped black string, the
thickness $r_0$ is only limited by the requirement that the ring remains
away from the nut. The ring thickness itself can be much larger than the
KK radius. This also remains valid with non-zero $b$, since a large
$B$-field $b\gg 1$ tends to make the coefficients $C_{\mu\nu}$ smaller.

\subsection{Physical parameters} 

We need to know how the parameters $r_0$ and $\alpha$ relate to the
number, mass and entropy of D0 branes. To find this we need the relation
between the Killing generators $\partial_\tau$, $\partial_z$ in the
region near the ring, and the canonical generators $\chi$ of the
asymptotic compact circles with period $2\pi L$, and $\zeta$ of
asymptotic time translations.

To this effect (see \cite{EEMR1}), we first note that the metric at
asymptotic infinity becomes
\beq
ds^2\to \frac{L^2}{4}\left(d\psi+(\cos\theta-1)d\phi -
\varpi d\bar t\right)^2 -d\bar t^2+dr^2+r^2d\theta^2+
r^2\sin^2\theta d\phi^2
\eeq
where $\bar t=2t/L$ is the canonically normalized time and
\beq
\varpi=\frac{2}{L}\frac{b(3-b^2)}{(1+b^2)^{3/2}}\,
\eeq
is the velocity $d\psi/d\bar t$ of the asymptotic frames. We now change
\beq
\psi=\frac{2}{L}y +\varpi\bar t+\bar\phi\,,\qquad \phi=\bar\phi
\eeq
in order to go to the canonical asymptotic form for the KK vacuum in its
rest frame,
\beq
ds^2\to \left(dy+\frac{L}{2}\cos\theta\; d\bar\phi\right)^2 -d\bar
t^2+dr^2+r^2d\theta^2+
r^2\sin^2\theta d\bar\phi^2
\eeq
with $y\sim y+2\pi L$.

The timelike Killing generator, $\zeta$, of the orbits of static
asymptotic observers is\footnote{Note that $\zeta$ differs from the
timelike Killing vector $\partial_t$ that is parallel to the
supersymmetry generators (\ie constructed as a bilinear of Killing
spinors).}
\beq\label{zetatpsi}
\zeta=\frac{\partial}{\partial{\bar t}}=
\frac{L}{2}\frac{\partial}{\partial t}+
\varpi\frac{\partial}{\partial\psi}\,,
\eeq
the generator $\chi$ of the Kaluza-Klein circle is
\beq\label{chipsi}
\chi=\frac{\partial}{\partial y}=
\frac{2}{L}\frac{\partial}{\partial\psi}\,,
\eeq
and the angular rotations along $\bar\phi$ are generated by
\beq\label{barphi}
\frac{\partial}{\partial \bar\phi}=
\frac{\partial}{\partial\phi}+\frac{\partial}{\partial\psi}\,.
\eeq

Now, since eq.~\eqref{chKill} gives
\beq
\frac{\partial}{\partial\psi}=\frac{\Delta z}{4\pi}\frac{\partial}{\partial z}\,,
\qquad
\frac{\partial}{\partial t}=\frac{4\pi}{\Delta
z}\frac{1}{\sqrt{H_R Z_R}}\left(\frac{\partial}{\partial\tau}-
\sqrt{\frac{H_R}{Z_R^3}}\omega_R\frac{\partial}{\partial z}\right)\,,
\eeq
then together with \eqref{zetatpsi} and \eqref{chipsi} we obtain the
relations we sought.

We can now compute the physical magnitudes following the analysis in
sec.~\ref{sec:physmag}. We shall express them as four-dimensional
quantities, taking into account that
\beq
G_4=\frac{G_5}{2\pi L}\,.
\eeq

The magnetic charge comes entirely from the background D6 and is
\beq
P=\frac{L}{4}\,,
\eeq
and 
the 4D electric charge is proportional to the momentum
along the compact direction,
\beq\label{Qel}
Q=2G_4\int dz d^3 x T_{\tau\mu}\chi^\mu=2G_4\frac{N_0}{L}
\eeq
where $N_0$ is given in \eqref{quantN0}. 

Since the black string has no angular momentum along $\phi$, it follows
from \eqref{barphi} that the four-dimensional angular momentum
associated to $\partial_{\bar\phi}$ is given by the Dirac value
\beq
J=\frac{Q P}{G_4}\,.
\eeq
If we had we taken the black string to be a Kerr black string then
$J\neq\frac{Q P}{G_4}$ --- this is illustrated in appendix
\ref{app:doublespin}. This is only possible when the D0 is excited, since in the limit $\alpha\to\infty$ the $S^2$ rotation must vanish to avoid pathologies, and so the Dirac relation must be satisfied.

The energy conjugate to time
translations generated by $\zeta$ is
\beq\label{masszeta}
M=\int dz d^3 x T_{\tau\mu}\zeta^\mu=
\frac{Q}{2 G_4}
\left[\left(\frac{2\pi L}{\Delta z}\right)^2\left(\frac{1}{\sqrt{H_R
Z_R}}\frac{\cosh^2\alpha+1}{\sinh\alpha\cosh\alpha}-
\frac{\omega_R}{Z_R^2}\right) +\frac{L\varpi}{2}\right]\,.
\eeq

The four-dimensional Einstein-frame area is
\beq
\mathcal{A}_4=\frac{\Delta z}{2\pi L}4\pi r_0^2\cosh\alpha\,,
\eeq
consistently with the invariance of the entropy under dimensional
reduction.
When the ring is non-extremal and $\alpha$ is finite, we can express the
area for a given electric charge (\ie the entropy for a given net number
of D0s) as
\beq\label{adq2}
\frac{\mathcal{A}_4}{16\pi Q^2}=
\frac{1}{\sinh^2\alpha\cosh\alpha}\left(\frac{2\pi L}{\Delta z}\right)^3\,,
\eeq
where $\Delta z$ is given in \eqref{Delz}.

\subsection{Equilibrium configurations}

Equilibrium configurations correspond to solutions of \eqref{noforce}.
For a black ring
\eqref{tensorcomponents} this requires
\beq\label{equil}
(\cosh^2\al+1)C_{\tau\tau}+2C_{\tau
z}\cosh\al\sinh\al=C_{zz}(\sinh^2\al-1)
\eeq
with $C_{\mu\nu}$
given by \eqref{cmunu}.
There are two simple situations of particular interest:

\subsubsection*{Extremal ring with $B\neq 0$}

When the ring is boosted to the speed of light, $\alpha\to\infty$ ---so
it is extremal and
chiral, corresponding to a D0 brane in its ground state--- the 
equilibrium equation \eqref{equilight} is solved for
\beq\label{equilR}
R=\frac{L}{2}\frac{(1+b^2)^{3/2}}{3b^2-1}.
\eeq
Thus, equilibria between D0 and D6 branes are possible for 
\beq\label{bcrit}
b>b_c=\frac{1}{\sqrt{3}}\,.
\eeq
In appendix~\ref{app:exactsusyd0d6} we compare the result \eqref{equilR}
with the one obtained from the exact supergravity solution, and show
perfect agreement when the D0 branes are treated perturbatively. The
critical value of the field \eqref{bcrit} also agrees with the value
computed in perturbative string theory \cite{EW1,Chen}.

\subsubsection*{Non-extremal ring with $B=0$}

When the D0 branes are thermally excited the boost $\alpha$ is finite. In
this case we can look for equilibrium configurations even when no $B$
flux is present, $b=0$. This simplifies greatly the background, since
$C_{\tau\tau}=C_{\tau z}=0$ and the equilibrium condition fixes the
boost value to
\beq\label{eqboost}
\sinh^2\alpha=1\,.
\eeq
In this case, we can write \eqref{adq2} 
(with $b=0$ in \eqref{Delz}) as
\beq
R=\frac{L}{2}\bigg[\bigg(\frac{\mathcal{A}_4}{8\sqrt{2}\pi Q^2}\bigg)^{2/3}-1\bigg]^{-1}
\eeq
which implies that, for a fixed net number of D0 branes (fixed $Q$), a bound
state can exist if the thermal excitation is large enough to create a horizon of area
\beq\label{acrit}
\mathcal{A}_4>\mathcal{A}_{4,c}=8\sqrt{2}\pi Q^2\,,
\eeq
or in terms of entropy and D0 number,
\beq\label{Scrit}
S>S_c=8\sqrt{2}\pi G_4\frac{N_0^2}{L^2}\,.
\eeq
It is important to observe that, even if we have derived this result
using the perturbative method for thin rings, the result \eqref{Scrit}
is actually {\it exactly} valid, since it is the value of the entropy
for a black ring bound at an infinite distance from the nut, in which
case \eqref{thinvalid} does not impose any constraint on the ring
thickness.

For $b=0$ and equilibrium boost \eqref{eqboost}, the expressions for the
physical parameters
take simple forms,
\beq\label{thinMAQ}
\frac{G_4 M}{Q}=\frac{3}{2\sqrt{2}}\left(\frac{2\pi L}{\Delta z}\right)\,,\qquad
\frac{\mathcal{A}_4}{(G_4 M)^2}=\frac{64\sqrt{2}\pi}{9}\left(\frac{2\pi L}{\Delta z}\right)\,.
\eeq
Since $\Delta z\leq 2\pi L$, with saturation when the
separation goes to infinity, we see that
not only the area, but
also the mass has a lower limit
\beq\label{Mcrit}
M>M_c=\frac{3}{2\sqrt{2}}\frac{Q}{G_4}=\frac{3}{\sqrt{2}}\frac{N_0}{L}
\eeq
for the bound state to exist.

\medskip

Thus we have demonstrated the two main mechanisms that permit the
formation of bound states of D0 and D6 branes. The general case in which
the two are simultaneously at work, \ie when both $b$ and $r_0$ are
finite, is only technically more difficult and not more illuminating, so
we will not dwell on it.

\subsection{Forces, interaction energy, and stability}
 
In a general configuration away from equilibrium, an external force
${\cal F}$ is needed in order to keep the D0 branes in place, which acts
at each point along the ring. Thus the ring exerts a force $-{\cal F}$,
and we can assign to it a potential energy
$dV_\mathcal{F}=\sqrt{-g_{tt}}\;\Delta
z\;\mathcal{F}$, redshifted from the location of the ring to asymptotic
infinity. Thus we introduce
\beq
V_\mathcal{F}(R)= -\int_R^\infty  \sqrt{-g_{tt}}\;\Delta z\;\mathcal{F}\,.
\eeq
Clearly, equilibrium corresponds to $V_\mathcal{F}'=0$. 

There is another measure of the energy of the interaction, given by
shift in the {\it internal} energy of the D0-branes when placed in the
field created by the D6 brane---the D6 is not affected since we are
regarding it as a background. This shift is the difference between the
measured mass $M$ of the D0s in the presence of the D6 and their mass in
isolation,
\beq
\Delta M(R) =M(R)-M_\mathrm{D0}\,.
\eeq
By the mass of isolated D0s we mean the mass of a D0-charged black hole
of given charge and area. Then $\Delta M$ measures the change in its
internal energy as it is moved adiabatically from infinity to $R$. When
the D0 is extremal, its mass in isolation is simply $Q/(2G_4)$.
We mentioned above that the mass is extremized for configurations that
satisfy the no-force condition. Thus, the extrema of $\Delta M$ coincide
with the extrema of $V_\mathcal{F}$. 

In principle $\Delta M$ and $V_\mathcal{F}$ seem to be different
contributions to the total interaction energy, and we might expect that
the quantity that corresponds to $E_\mathrm{int}$ in \eqref{eintdef},
which we use for exact solutions, is the sum of both. In fact we have
checked explicitly that, at large distances, the value of
$E_\mathrm{int}$ for the exact solutions coincides with that of $\Delta
M+V_\mathcal{F}$.

On the other hand, not only do the minima of $\Delta M$ and
$V_\mathcal{F}$ coincide, but the two functions also resemble each other
closely for generic $b$ and $r_0$, and in fact agree exactly for $b=0$. Whether adding up $\Delta M$ and $V_\mathcal{F}$ is the
correct procedure, or instead is double-counting the interaction energy,
is not completely clear to us,\footnote{The fact that this prescription
would give a larger total mass to the system at equilibrium than
$M_\mathrm{D6}+M$ points in this direction.} but fortunately none of
our conclusions depends on this, since the properties of the interaction
energy remain the same (up to possibly a factor close to 2) with either
definition.
The interaction energy $E_\mathrm{int}=\Delta
M+V_\mathcal{F}$ for the two particular cases of interest
discussed above has been presented in figs.~\ref{fig:D0D6B} and
\ref{fig:D0D6therm}. Here we discuss some of their properties at short
and long distances.

If we consider first $R\to 0$ it is easy to see that in the extremal
case at fixed charge, the mass $M$ in \eqref{masszeta} diverges as
$1/\sqrt{R}$, and so does then, too, $\Delta M$.
Thus at short distances we cannot expect our perturbative approximation
to remain valid and we must resort to solutions that account for
backreaction of the D0. This is addressed in sec.~\ref{sec:exactextr}.

At large distances the values of $\Delta M$ and
$V_\mathcal{F}$ are equal to leading order. Adding them together to obtain the
total interaction energy we find
\beq
G_4 E_\mathrm{int}(R)=-\left(2 b^3 +
\sqrt{1+b^2}\frac{(2 b^2-1)\sinh^2\alpha_\infty
+b^2+1}
{\cosh\alpha_\infty\sinh\alpha_\infty}\right) \frac{Q
P}{R}+O(R^{-2})\,,
\eeq
where $\alpha_\infty$ is a function of $Q^2/\mathcal{A}_4$, independent of $b$,
determined as the solution of 
\eqref{adq2} at $R\rightarrow\infty$, \ie $\Delta z\to 2\pi L$,
\beq\label{adqinf}
\sinh^2\alpha_\infty\cosh\alpha_\infty=\frac{16\pi Q^2}{\mathcal{A}_4}\,.
\eeq

Depending on the sign of the coefficient of $QP/R$ the interaction
energy will be repulsive or attractive. This is in fact in line with the
force analysis, since this coefficient is proportional to the force
$\mathcal{F}$ at $R\to\infty$, and when it vanishes a bound state at
infinity appears. We can see this more explicitly in the two particular
cases of interest.
For the extremal case, $\alpha_\infty\to\infty$ ($\mathcal{A}_4\to 0$) the above
expression simplifies to
\beq
G_4 E_\mathrm{int}(r)=
\frac{1-3b^2}{2 b^3-\sqrt{1+b^2}(2b^2-1)}
\frac{Q
P}{R}+O(R^{-2})\,.
\eeq
This interaction energy
changes from positive, hence
repulsive, for $b<1/\sqrt{3}$ to negative, hence attractive,
for $b>1/\sqrt{3}$.  The critical value $b_c=1/\sqrt{3}$ agrees
with the expectation that the long-distance perturbative string
interaction changes sign at $b=b_c$.
At $b=0$ it becomes particularly simple,
\beq\label{eintb=0}
G_4 E_\mathrm{int}(R)=\frac{QP}{R}+O(R^{-2})\,.
\eeq

For the nonextremal ring with $b=0$ we have
\beq
G_4
E_\mathrm{int}(r)=\frac{\sinh^2\alpha_\infty-1}{\cosh\alpha_\infty
\sinh\alpha_\infty}\frac{Q
P}{R}+O(R^{-2})\,.
\eeq
We see that bound a state at infinity appears for
$\sinh^2\alpha_\infty=1$ as expected, and that the interaction becomes
attractive when the ring at infinity is rotating more slowly,
$\sinh^2\alpha_\infty<1$: the centrifugal force at infinity is weaker so
the ring will tend to shrink and move towards the nut.

\section{Exact solution-generating technique }
\label{constructionsec}
In the case of five-dimensional stationary vacuum solutions with two $U(1)$ isometries, one can actually go beyond the perturbative approximation and construct exact geometries by the application of a combination of solution-generating techniques. In particular, we will use the method of \cite{GS1} and \cite{GS2} to generate an asymptotically Taub-NUT solution starting from an asymptotically flat solution. This method is based on the application of an appropriate element of the $SL(3,\mathbb R)$ group of symmetries, discovered in \cite{Maison}, that the solutions to the Einstein vacuum equations with these isometries have.  In this section, after briefly recalling the solution generating technique that we employ, we will show how to construct an exact solution corresponding to a black ring in Taub-NUT.

\subsection{Review of the solution-generating method}

The starting solution, that will be referred to  as the ``seed'', is a stationary axisymmetric solution (\cite{TH}) in five dimensions:
\be
ds^2= G_{IJ} dy^I dy^J + e^{2\nu} (d\rho^2+dz^2)\,,
\label{seedone}
\ee 
where $y^I$, $I=0,1,2$,  are coordinates corresponding to Killing directions of the solution; in our case $y^I=\{t,\hat{\phi},\hat{\psi}\}$, with $\hat{\phi}$ and $\hat{\psi}$ the Cartan angles of $\mathbb{R}^4$ . Here the metric coefficients $G_{IJ}$ and $e^{2\nu}$ only depend on the Weyl coordinates $\rho$ and $z$. The $SL(3,\mathbb R)$ transformation that relates asymptotically flat and Taub-NUT solutions acts naturally on the
Euler angles $\phi_\pm$:
 \begin{equation}
\hat{\psi}=\frac{1}{2}(\phi_++\phi_-)\,,~~~~~~\hat{\phi}=\frac{1}{2}(\phi_+-\phi_-)\,. \label{eqn:newcoords}
\end{equation}
We will identify the fiber of Taub-NUT space with the direction $\phi_+$. Introducing 
coordinates $\xi^0\equiv t$ and $\xi^1=\ell \phi_+$, where $\ell$ is an arbitrary length scale, it is useful to  rewrite the metric (\ref{seedone}) in the form
\begin{equation}
ds^2=\lambda_{ab}(d\xi^a+\omega^a)(d\xi^b+\omega^b)+\frac{1}{\tau}\,ds^2_3\,,
\label{seedmaison}
\end{equation}
where $a,b=0,1$, $\tau=-\det \lambda_{ab}$ and $\omega^a=\omega^a_- d\phi_-$ are one-forms on the base space $ds^2_3$.   The three-dimensional metric $ds^2_3$ on the base space is then given by
\be
ds^2_3 = \tau \,e^{2\nu} (d\rho^2+dz^2) + {\rho^2\over 4\,\tau}\, d\phi_-^2\,.
\label{baseseed}
\ee
Using the fact that in three dimensions a one-form is dual to a scalar, we can  introduce the potentials $V_a$:
\be
d V_a = -\tau \lambda_{ab} *_3 d\omega^b\,,
\label{duality}
\ee
where the Hodge operation $*_3$ is performed with the metric $ds^2_3$. It can be shown that the integrability condition of this equation is satisfied thanks to the Einstein equations.   Then, the data contained in the metric (\ref{seedmaison}) can be re-packadged into the symmetric unimodular matrix of scalars
$\chi$:
\begin{equation}
 \chi = \left(
\begin{array}{cc}
 \lambda_{ab}-\frac{V_aV_b}{\tau} & \frac{V_a}{\tau} \\
 \frac{V_b}{\tau} & -\frac{1}{\tau}
\end{array}
\right)\,. \label{eqn:chimat}
\end{equation}
The usefulness of this formalism relies on the fact that the equations of motion are left invariant
by the action of an $SL(3,\mathbb R)$ group of transformations that act linearly on $\chi$:
\be
\chi\to \chi'=N \,\chi \,N^T\;,\qquad ds^2_3\to ds^2_3\;,\qquad N\in SL(3,\mathbb R)\;.
\ee
This provides a solution generating method: starting from a solution $(\chi, ds_3^2)$, one can construct a new solution $(\chi',ds^2_3)$ by acting on the former with  suitable elements of $SL(3,\mathbb R)$. Reconstruting the final metric from the rotated matrix $\chi'$ requires inverting the duality relations (\ref{duality}) in order to compute the transformed one-forms $\omega'^a$. As a computational trick to facilitate this procedure, it is useful to introduce a matrix of one-forms $\kappa$, defined as
\be
*_3 d\kappa =\chi^{-1} d\chi \,.
\ee
One can show that $\kappa$ encodes the information about the one-forms $\omega^a$ since
\be
\omega^a=-\kappa^a_{\phantom a 2} \quad (a=0,1)\,.
\ee
Moreover, the definition of $\kappa$ implies that it also transforms linearly under $SL(3,\mathbb R)$:
 \be
 \kappa\to\kappa'=(N^{-1})^T\,\kappa\,N^T\;,\qquad N\in SL(3,\mathbb R)\;.
 \ee
Therefore, by following the transformations of both $\chi$ and $\kappa$ under  $SL(3,\mathbb R)$, one can easily reconstruct the transformed metric by purely algebraic manipulations. We will refer to the 
 set of data $(\chi,\kappa)$ as the Maison data.
 
In \cite{GS1}, \cite{GS2} the element of  $SL(3,\mathbb R)$ that maps five-dimensional asymptotically flat solutions into asymptotically Taub-NUT solutions was identified. To construct a black ring in Taub-NUT, one should, naively, apply this $SL(3,\mathbb R)$ transformation to the black ring  of \cite{ER1}. However, as explained in \cite{GS2}, this does not quite work: the  $SL(3,\mathbb R)$ transformations change the relative orientation of the rods  which can spoil the regularity of the solution. The application of an $SL(3,\mathbb R)$ transformation to a solution with a regular horizon of  topology  $S^2\times S^1$ produces, in general, a solution with a singular horizon. To counterbalance this effect, one should start from a singular seed solution that generalizes the flat space black ring and contains an extra parameter encoding the relative orientation of the space-like rods on either side of the horizon. This extra parameter is then fixed in such a way that the $SL(3,\mathbb R)$-transformed solution has a regular ring-like horizon.  

The easiest way to produce the needed seed solution is to use  the BZ \cite{BZ1} (see \cite{BV} for a detailed review) technique to construct a one-parameter family of solutions that generalizes the black ring of \cite{ER1}. Given a solution generated by the BZ method, a series of technical
results derived in \cite{GS2}  allow to compute the corresponding $\chi$ and $\kappa$ matrices. In section \ref{seedsec} we will construct the appropriate seed and
compute the associated Maison data. The $SL(3,\mathbb R)$ transformation will be carried out in section  \ref{rodsec}, where we will also perform a singularity analysis and determine the values of parameters for which the solution is regular. 

\subsection{Seed solution}
\label{seedsec}
In this section we construct the appropriate seed solution to which we will apply a suitable $SL(3,\mathbb R)$ transformation that will eventually yield a black ring in Taub-NUT. The data encoding the seed solution are the metric factors $\{G_{IJ},\,e^{2\nu}\}$ defined in (\ref{seedone}).

\subsubsection*{The metric}
Our starting point is a Weyl solution  given by
\begin{equation}
 G_0=\textrm{diag}\left\{-\frac{\mu_1}{\mu_3},\frac{\rho^2\mu_3}{\mu_2\mu_4},\frac{\mu_2\mu_4}{\mu_1}\right\}\,.
\label{eqn:seed}
\end{equation}
The first term in $G_0$ corresponds to the $tt$-component, the second to the $\hat{\phi}\hat{\phi}$-component and the third to the $\hat{\psi}\hat{\psi}$-component. In additon, we  have
\begin{equation}
 e^{2\nu_0}=k^2~
\frac{\mu_2\mu_4(\rho^2+\mu_1\mu_2)(\rho^2+\mu_1\mu_3)(\rho^2+\mu_1\mu_4)(\rho^2+\mu_2\mu_3)(\rho^2+\mu_3\mu_4)}{\mu_1(\rho^2+\mu_2\mu_4)^2\prod_{i=1}^4(\rho^2+\mu_i^2)}\,,
\label{eqn:cfseed}
\end{equation}
with $k>0$ without loss of generality. We use the standard notation
\be
\mu_i = \sqrt{\rho^2 + (z-a_i)^2}-(z-a_i)\,,\qquad \bar\mu_i=-\rho^2/\mu_i\,,\qquad i=1\ldots 4
\ee
and we assume the ordering 
\begin{equation}
 a_1\leq a_2 \leq a_3 < a_4\,,
\end{equation}
of the rod endpoints. The initial solution \eqref{eqn:seed}-\eqref{eqn:cfseed} is singular by itself, but as explained in \cite{saturn}, the singularity can be canceled after the soliton transformations by fixing the BZ parameters conveniently.

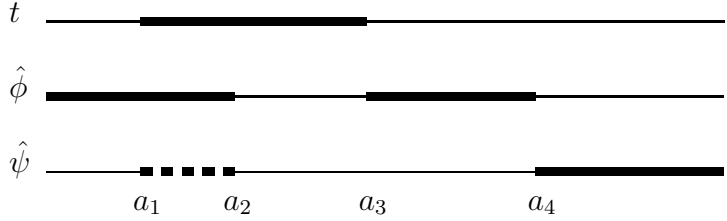
\begin{figure}[t]
\vspace{2cm}
\begin{center}
\begin{picture}(0,0)
\setlength{\unitlength}{1cm}
\put(-5,1.5){\line(1,0){2.5}}
\put(-3.75,1.5){\linethickness{0.1cm}{\line(1,0){3}}}
\put(-0.75,1.5){\line(1,0){4.75}}
\put(1.5,0.5){\line(1,0){2.5}}
\put(-5,0.5){\linethickness{0.1cm}{\line(1,0){2.5}}}
\put(-2.5,0.5){\line(1,0){1.75}}
\put(-0.75,0.5){\linethickness{0.1cm}{\line(1,0){2.25}}}
\put(-5,-0.5){\line(1,0){1.25}}
\put(-2.5,-0.5){\line(1,0){5}}
\put(1.5,-0.5){\linethickness{0.1cm}{\line(1,0){2.5}}}
\multiput(-3.75,-0.5)(0.275,0){5}{\linethickness{0.1cm}\line(1,0){0.15}}
\put(-5.5,1.5){$t$}
\put(-5.5,0.5){$\hat{\phi}$}
\put(-5.5,-0.5){$\hat{\psi}$}
\put(-3.85,-1){$a_1$}
\put(-2.65,-1){$a_2$}
\put(-0.85,-1){$a_3$}
\put(1.4,-1){$a_4$}
\end{picture}
\end{center}
\vspace{0.5cm}
\caption{\small{Rod structure of the initial solution \eqref{eqn:seed}. This is the same rod structure of the seed solution of the standard $S^1$--spinning ring.}}
\label{fig:seed}
\end{figure}

We generate the wanted seed solution  by means of a two-soliton transformation on \eqref{eqn:seed}. We briefly summarize the steps of this construction:
\begin{enumerate}
 \item Remove an anti-soliton at $z=a_1$ and a soliton at $z=a_4$ form $(G_0)_{tt}$ and $(G_0)_{\hat{\phi}\hat{\phi}}$ respectively. The resulting metric is:
\begin{equation}
 G_1=\textrm{diag}\left\{\frac{\rho^2}{\mu_1\mu_3},-\frac{\mu_3\mu_4}{\mu_2},\frac{\mu_2\mu_4}{\mu_1}\right\}\,.
\end{equation}
\item Rescale the metric by a factor of $1/\mu_4$ to find
\begin{equation}
 \tilde G_0=\frac{1}{\mu_4}\,G_1=
\textrm{diag}\left\{-\frac{\bar\mu_3}{\mu_1\mu_4},-\frac{\mu_3}{\mu_2},\frac{\mu_2}{\mu_1}\right\}\,.
\end{equation}
This is the new seed solution to which we apply the BZ transformations. The corresponding generating matrix is then given by
\begin{equation}
 \Psi_0=\textrm{diag}\left\{
	-\frac{(\bar\mu_3-\lambda)}{(\mu_1-\lambda)(\mu_4-\lambda)},
	-\frac{(\mu_3-\lambda)}{(\mu_2-\lambda))},
	\frac{(\mu_2-\lambda)}{(\mu_1-\lambda)}\right\}\,.
\end{equation}
\item Perform now a two-soliton transformation with $\tilde G_0$ as seed, re-adding the anti-soliton at $z=a_1$ and the soliton at $z=a_4$ with BZ vectors $m_{0}^{(1)}=(1,0,b_1)$, $m_0^{(4)}=m(0,1,b_4)$ respectively. Denote the resulting solution by $\tilde G$.
\item Rescale $\tilde G$ to find the final metric:
\begin{equation}
 G=\mu_4\,\tilde G\,.\label{eqn:finalG}
\end{equation}
Note that by construction $G$ satisfies $\det G=-\rho^2$.
\end{enumerate} 

The solution obtained at this stage has a naked singularity at $z=a_1$, the position of the negative density rod of the starting metric $G_0$. This singularity can be removed by fixing the BZ parameter $b_1$ to be
\begin{equation}
 |b_1|=\sqrt{\frac{2(a_2-a_1)(a_4-a_1)}{(a_3-a_1)}}\,. \label{eqn:b1}
\end{equation}
From now on we will only consider the solution with $b_1$ fixed as above. The sign of $b_1$ is arbitrary and it can always be changed by changing $\hat{\psi}\to-\hat{\psi}$, which corresponds to inverting the sense of rotation of the five-dimensional asymptotically flat solution. Without loss of generality, we choose $b_1>0$.

\subsubsection*{Parametrization}
Having fixed $b_1$, the resulting solution can be conveniently parametrized so that its relation with the five-dimensional asymptotically flat black ring is manifest. Following \cite{saturn}, we choose
\begin{equation}
 a_1=-R^2\,\frac{2\lambda-\nu(1+\lambda)}{2(1-\lambda)}\,,~~~~~~a_2=-\frac{R^2}{2}\nu\,~~~~~~a_3=\frac{R^2}{2}\nu\,,~~~~~~a_4=\frac{R^2}{2}\,.
\end{equation}
Notice that in this paramtrization $b_1$ is given by
\begin{equation}
 b_1=R\sqrt{\frac{(1+\lambda)(\lambda-\nu)}{\lambda(1-\lambda)}}\,.
\end{equation}
We find it convenient to change to the C-metric type of coordinates of \cite{RE1}, for which
\begin{equation}
 \rho=\frac{R^2}{(x-y)^2}\sqrt{-G(x)G(y)}\,,~~~~~~z=\frac{R^2(1-xy)\big[1+\nu(x+y)\big]}{(x-y)^2}\,,
\end{equation}
where $G(\xi)$ is  defined below.  Notice that, defining $R_i=\sqrt{\rho^2+a_i^2}$, $i=2,3,4$, we can invert the relations above and write $(x,y)$ in terms of $(\rho,z)$:
\begin{subequations}
\begin{align}
x&=\frac{(1-\nu)R_2-(1+\nu)R_3-2R_4+R^2(1-\nu^2)}{(1-\nu)R_2+(1+\nu)R_3+2\nu R_4}\;,\\
y&=\frac{(1-\nu)R_2-(1+\nu)R_3-2R_4-R^2(1-\nu^2)}{(1-\nu)R_2+(1+\nu)R_3+2\nu R_4}\;.
\end{align}
\end{subequations}

Finally, rescaling $b_4$ as $b_4=\bar b_4 (1+\nu)^2\sqrt{\frac{1-\lambda}{1+\lambda}}$, we can write the metric \eqref{eqn:finalG} in a simple looking form,\footnote{This metric is a particular case of the solution constructed in \cite{TEO}. However at this stage we have not imposed any condition on the parameters that determine the directions of the rods.}
\begin{equation}
\begin{aligned} 
ds^2=&-\frac{H(y,x)}{H(x,y)}\Big[dt+\Omega\Big]^2\\
	&\!\!+\!\frac{R^2}{(x-y)^2}\left[-\frac{F(y,x)}{H(y,x)}d\hat{\psi}^2+k^2\,H(x,y)\!\left(\!\!-\frac{dy^2}{G(y)}
	+\frac{dx^2}{G(x)}\!\right)\!+\frac{F(x,y)}{H(y,x)}d\hat{\phi}^2
	+\frac{2J(x,y)}{H(y,x)} d\hat{\phi} d\hat{\psi}\right],\label{eqn:metricxy}
\end{aligned}
\end{equation}
where
\begin{eqnarray}
G(x)&=&(1-x^2)(1+\nu x)\,,\\
H(x,y)&=&1+\lambda x-\bar b_4^2(1+\nu x)^2(1+\nu y)\big[1-\lambda\nu-(\lambda-\nu)y\big]\,,\\
F(x,y)&=&G(x)(1+\lambda y)-\bar b_4^2G(y)(1+\nu x)^3\big[1-\lambda\nu-(\lambda-\nu)x\big]\,,\\
J(x,y)&=&-(1+\lambda)\bar b_4\, C_2\,(x-y)(1+\nu x)(1+\nu y)\big[x+y+\nu(1+xy)\big]\,,
\end{eqnarray}
and the rotation one-form $\Omega$ is given by
\begin{equation}
 \Omega=\frac{R\,C_1}{H(y,x)}\big[\omega_{\hat{\psi}}(x,y)d\hat{\psi}+\bar b_4\, C_2\, \omega_{\hat{\phi}}(x,y) d\hat{\phi}\big]\,,
\end{equation}
with
\begin{eqnarray}
 \omega_{\hat{\psi}}(x,y)&=&1+y-\bar b_4^2(1-\nu)(1-x)(1+\nu x)(1+\nu y)^2\,,\\
\omega_{\hat{\phi}}(x,y)&=&(1+\nu x)\big[x+y+\nu(1+xy)\big]\,,
\end{eqnarray}
and
\begin{equation}
 C_1=\sqrt{\lambda(\lambda-\nu)\frac{1+\lambda}{1-\lambda}}\,,~~~~~~
 C_2=\sqrt{\frac{1-\lambda}{1+\lambda}}\,.
\end{equation}
Note that we have left the constant $k$ in front of the conformal factor unspecified. We will fix it later on when we consider the asymptotics of the final metric. Finally, the dimensionless parameters $\lambda$ and $\nu$ must lie in the range
\begin{equation}
 0< \nu \leq \lambda < 1\,.
\end{equation}

The metric  \eqref{eqn:metricxy} is not written in a manifestly asymptotically flat form. Though one could perform a change of coordinates to bring the metric in an asymptotically flat frame, this is not needed for
our construction. Furthermore, the solution \eqref{eqn:metricxy} has closed timelike curves due to the fact that the direction of the finite spacelike rod $z\in[a_3,a_4]$, has a component along  $t$, which implies that $t$ has to be globally identified with a certain period. In fact, this corresponds to the presence of a Dirac-Misner string. As explained above, this problem will be cured after the action of $SL(3,\mathbb R)$. Finally, we notice that setting $b_4=0$, or equivalently $\bar b_4=0$, the metric \eqref{eqn:metricxy} reduces to the $S^1$--spinning ring of \cite{RE1}.

To apply the $SL(3,\mathbb R)$ transformation that will generate the metric for a black ring in Taub-NUT, one needs to compute the Maison data $(\chi,\kappa)$ for the metric \eqref{eqn:metricxy}. This computation is rather involved and the interested reader can find it in appendix \ref{sec:appMaison}.

\subsection{Constructing the ring}
\label{rodsec}

To proceed with the construction of the solution, we  examine first the rod structure \cite{TH} of the seed solution  \eqref{eqn:metricxy}. There are the following four rods: rod 1, at $z\in (-\infty, -{R^2\over 2} \nu]$ (or $x=-1$); rod 2, at  $z\in [-{R^2\over 2}\nu, {R^2\over 2} \nu]$ (or $y=-{1\over\nu}$); rod 3, at $z\in [{R^2\over 2}\nu, {R^2\over 2}]$ (or $x=1$); rod 4, at $z\in [{R^2\over 2}, \infty)$ (or $y=-1$). As shown in \cite{GS1}, the eigenvectors, $v_i$ ($i=1,\ldots,4$), associated to each rod can be easily derived from the matrix $\kappa$ as
\be
v_i=\lim_{\rho\to 0} (\kappa_{02},\kappa_{12},1)\Big|_{z\in I_i}\,,
\label{rodeigen}
\ee
where $I_i$ is the interval corresponding to the $i$-th rod, and we are writing the vectors in the basis
$\{{\partial\over\partial t},{\partial\over\partial\xi^1}, {\partial\over\partial \phi_-}\}$. This result holds whenever $\tau\not =0$, which  is satisfied on every rod for our seed solution. 

The rod 2 is  timelike and therefore corresponds to the  horizon of the solution.  In order for the topology of this horizon be $S^2\times S^1$, it is necessary that its neighboring rods, namely rods 1 and 3, have the same direction, i.e. $v_1=v_3$. One can check, using (\ref{rodeigen}), that our seed solution does not satisfy this requirement and hence it is not a black ring.\footnote{Instead, this solution can be interpreted as black lens \cite{TEO}.} As explained above, however, this feature of the seed solution is exactly what is needed to produce a regular black ring in Taub-NUT: the application of a suitable $SL(3,\mathbb R)$ transformation will add KK-monopole charge
and, at the same time, modify the relative orientation of rods 1 and 3 in such a way that the final solution
will satisfy $v_1=v_3$. Let us see how this works in some detail.

To add KK-monopole charge to an asymptotically flat solution, one should apply to the seed solution the transformation $D$, where $D$ is a particular element of  $SL(3,\mathbb R)$ (see below) \cite{GS1}. However, the action of $D$ generates unwanted Dirac-Misner strings. This pathology can be canceled by further acting on the solution with an element in the $SO(2,1)$ subgroup of $SL(3,\mathbb R)$ that preserves the asymptotic boundary conditions. There are three such transformations, denoted as $N_\alpha$, $N_\beta$ and $N_\gamma$ in \cite{DR1}. $N_\alpha$ is equivalent to a ``boost'' in the $\xi^1$ direction. Such a transformation does not change the relative orientation of the rods, nor the periodicity of the angular coordinates and, for this reason, does not affect the topology of the horizon. 
It has been shown in \cite{GS1} that the action $N_\beta$ is equivalent to a redefinition of the scale $\ell$ and  hence, if one keeps $\ell$ as an arbitrary parameter, the action of $N_\beta$ is  superfluous. On the other hand, $N_\gamma$ changes the regularity properties of the geometry, and we will need to include its action to obtain the desired solution.

In a first step we obtain an asymptotically Taub-NUT solution by acting on the seed  \eqref{eqn:metricxy} with $D$ and $N_\gamma$. In terms of the Maison data $(\chi',\kappa')$, the new solution is given by
\be
\chi'=N_\gamma D\, \chi \,D^T N_\gamma^T \,,\quad \kappa' = N_\gamma D\, \kappa\, D^T N_\gamma^T\,,
\label{newchiandkappa}
\ee
where
\be
D=\begin{pmatrix}1&0&0\cr 0&{1\over \sqrt{2}}&{1\over \sqrt{2}}\cr 0&-{1\over \sqrt{2}}&{1\over \sqrt{2}}\end{pmatrix}\,,\quad N_\gamma= \begin{pmatrix}\cos\gamma&0&-\sin\gamma\cr 0&1&0\cr \sin\gamma&0&\cos\gamma\end{pmatrix}\,.
\ee
The resulting metric has rods at the same positions as the seed metric but with different orientations, which are determined by the parameters of the $SL(3,\mathbb R)$ transformation. 
The eigenvectors corresponding to the rods 1 and 3 of the solution in (\ref{newchiandkappa}) have the form
\bea
v'_1&=& \Bigl(\ell(c^0_1 \cos2\gamma+s^0_1 \sin2\gamma),\ell(c^1_1 \cos\gamma+s^1_1 \sin\gamma),1\Bigr)\,,\nonumber\\
v'_3&=&\Bigl(\ell(c^0_3 \cos2\gamma+s^0_3 \sin2\gamma + z^0_3) ,\ell(c^1_3 \cos\gamma+s^1_3 \sin\gamma),1\Bigr)\,,
\eea
where $c^i_j$, $s^i_j$ and $z^0_3$ are some functions of $b_4$, $\lambda$, $\nu$ and the dimensionless ratio
\be
\hat{R}\equiv {R\over\ell}\,.
\ee 
We should require that $v'_1=v'_3$ in order for the new solution be a black ring. The $\xi^1$-component of this equation fixes the angle $\gamma$ in $N_\gamma$ as
\be
\tan\gamma = -{c^1_1-c^1_3\over s^1_1-s^1_3}\,.
\label{gammasol}
\ee
Imposing the equality of the $t$-components of $v'_1$ and $v'_3$, and using the
value of $\gamma$ found above, leads to the condition
\be
(c^0_1-c^0_3)[(s^1_1-s^1_3)^2-(c^1_1-c^1_3)^2]-2(s^0_1-s^0_3)(c^1_1-c^1_3)(s^1_1-s^1_3)-z^0_3[(s^1_1-s^1_3)^2+(c^1_1-c^1_3)^2]=0 \,,
\label{beq}
\ee
which is an algebraic equation for $b_4$. Solving this equation, and substituting the value of $b_4$ in (\ref{gammasol}), leaves as free parameters $\ell$, $R$, $\nu$ and $\lambda$, which are the parameters one expects for a black a ring in Taub-NUT with one independent angular momentum: $\ell$ sets the scale of the KK circle, $R$ is a measure of the radius of the ring, $\nu$ is a measure of the ratio between the radii of $S^2$ and $S^1$ at the horizon, and $\lambda$ controls the angular momentum in the plane of the ring. 

\subsection{Balanced rings}
In order to avoid a conical singularity at the location of a given rod, the period $\Delta_i$ of the spacelike coordinate $\phi_i(=$ a linear combination of $\phi_+,\phi_-)$ vanishing there must be fixed as
\be
\Delta_i= 2\pi\lim_{\rho\to 0}\sqrt{\rho^2 g'_{\rho\rho}\over |v'_i|^2}\qquad \textrm{for}\qquad z\in I_i\,,
\ee
where $g'_{\rho\rho}$ is the $\rho\rho$ component of the metric determined by the Maison data (\ref{newchiandkappa}), and $|v'_i|$ is the norm of $v'_i$. We find,
\begin{subequations}
\begin{align}
 \Delta_1&=\Delta_4=2\pi\,\frac{k\,\big|b_4(1+\lambda)(1-\nu)^2+(1-\lambda)(1+\nu)^2\big|}{(1-\lambda)(1+\nu)^2}\;,\\
\Delta_3&=2\pi\,\frac{k\,|1+b_4|(1-\nu)}{1+\nu}\sqrt{\frac{1+\lambda}{1-\lambda}}\;.
\end{align}
\end{subequations}
Since for a ring one has $v'_1=v'_3$, one needs to impose $\Delta_1=\Delta_3$ in order to cancel the possible conical singularities. This condition can be solved for $\lambda$, and one finds that there are two solutions
\begin{eqnarray}
\lambda&=&{2\nu\over 1+\nu^2}\,,
\label{lambdasol}\\
\lambda&=&{(1+\nu)^2-b_4^2(1-\nu)^2\over (1+\nu)^2+b_4^2(1-\nu)^2}\,.
\label{secondlambda}
\end{eqnarray}
It can be checked that for the value of $\lambda$ given in (\ref{secondlambda}) 
the rod structure of the solution degenerates:
the eigenvector of the semi-infinite rod, $v'_1$, becomes parallel to the eigenvector of the other semi-infinite rod, $v'_4$. This is the rod structure of a solution which has $\mathbb{R}^{3,1}\times S^1$ asymptotics, rather than Taub-NUT, and we can thus discard this solution.
The requirement of absence of conical singularities, then uniquely fixes $\lambda$ to take the value (\ref{lambdasol}), which is the same value one finds for the $S^1$-rotating black ring in flat space. Notice this coincides with the perturbative analysis of section \ref{sec:pertd0d6}. This is the case we will consider in the following. 

Restricting to balanced rings also simplifies the expressions considerably. Choosing,\footnote{This guarantees that the four-dimensional radial coordinate is canonically normalized.}
\begin{equation}
k=\frac{2}{|1+b_4|} \label{eqn:thek}
\end{equation}
the periodicities of the angular directions become
\begin{equation}
\Delta\phi_-=\Delta\phi_+=4\pi\;, \label{eqn:period5d}
\end{equation}
and the value of the angle $\gamma$ is then given by
\be
\tan\gamma= \hat{R}\,\nu\,\sqrt{1+\nu\over 1-\nu}\, {2(1-\nu) +\hat{R}^2 -b_4 (2(1-\nu) -\hat{R}^2)\over (1-\nu)+\hat{R}^2 (1+2\nu-\nu^2) +b_4 \hat{R}^2 (1+\nu)^2} \,.
\label{gammasolbis}
\ee

As for the parameter $b_4$, one finds that equation (\ref{beq}) admits the following three solutions:
\bea
b_4^{(1)}&=&-{4(1-\nu)+\hat{R}^2 (1+\nu^2) \over \hat{R}^2 (1-\nu^2)}\,,\\
b_4^{(2)}&=&{1\over \hat{R}^2 (4-\hat{R}^2) (1-\nu^2)^2}\Bigl[2 \sqrt{(2(1-\nu)+\hat{R}^2 \nu)^{3} (2(1-\nu) -\hat{R}^2 \nu (1-2\nu^2))} \nonumber\\
&&\qquad\qquad\qquad\quad - 8(1-\nu)^2 -4 \hat{R}^2 \nu(1-\nu)(1+\nu^2) +\hat{R}^4 (1-2\nu^2-\nu^4)\Bigr]\,, \label{bsol}\\
b_4^{(3)}&=&{1\over \hat{R}^2 (4-\hat{R}^2) (1-\nu^2)^2}\Bigl[-2 \sqrt{(2(1-\nu)+\hat{R}^2 \nu)^{3} (2(1-\nu) -\hat{R}^2 \nu (1-2\nu^2))} \nonumber\\
&&\qquad\qquad\qquad\quad - 8(1-\nu)^2 -4 \hat{R}^2 \nu(1-\nu)(1+\nu^2) +\hat{R}^4 (1-2\nu^2-\nu^4)\Bigr]\,.
\eea
Out these, only $b_4^{(2)}$ leads to a regular black ring. One finds, indeed, that the metric corresponding to $b_4^{(1)}$ has a degenerate rod structure. The rod 2  has the same direction as its two neighboring (spacelike) rods and  the full rod structure is identical to that of flat space.  Moreover the metric has singularities at the end-points of rod 2. For the solution $b_4^{(3)}$, we have checked numerically that the corresponding final metric has both naked singularities and CTCs outside the horizon, and it is thus physically unacceptable. This leaves us with only the solution $b_4^{(2)}$. One can verify that the corresponding metric is regular and free of CTCs, and, moreover, reduces to the $S^1$-rotating black ring in flat space when the radius of the KK direction becomes much larger than the scale of the ring. Therefore, from now on we will only consider the solution with $b_4=b_4^{(2)}$.

\subsection{Final solution}

At this stage, our solution for the black ring in Taub-NUT is specified by  $b_4^{(2)}$, $\gamma$ and $\lambda$ given in equations \eqref{bsol}, \eqref{gammasolbis} and \eqref{lambdasol} respectively.   The geometry corresponding to the data (\ref{newchiandkappa}) has a horizon with topology $S^2\times S^1$ and no conical singularities.  However, the solution has Dirac-Misner strings, which can be seen from the fact that the directions of the spacelike rods have compoenents along $\partial_t$. To cure this pathology  one still needs to apply a transformation $N_\alpha\in SO(2,1)$: this leads to a metric specified by the following Maison data
\be
\chi''=N_\alpha\, \chi' \, N_\alpha^T\,,\quad \kappa''=(N_\alpha^{-1})^T \,\kappa'\, N_\alpha^T\,,
\label{finalchiandkappa}
\ee
where
\be
N_\alpha= \begin{pmatrix}\cosh\alpha&\sinh\alpha&0\cr \sinh\alpha&\cosh\alpha&0\cr 0&0&1\end{pmatrix}\,.
\ee
 The value of $\alpha$ needed to cancel the Dirac-Misner strings is found to be 
\begin{equation}
\begin{aligned}
\tanh\alpha = \frac{\hat R\, \nu(1-\nu)D_\nu D_2-(1+\nu)^2D_1\,s_{2\gamma}}
{-4(1+\nu)^2D_3\,c_\gamma+2\,\hat R\,\nu(1-\nu)D_\nu\big[D_2+8\,b_4(1-\nu)\big]s_\gamma}\,,
\end{aligned}
\label{alphasol}
\end{equation}
where $s_\gamma\equiv \sin\gamma$, $c_\gamma\equiv \cos\gamma$ and the constants $D_\nu$, $D_1$, $D_2$ and $D_3$ are defined in appendix \ref{app:charges}. Therefore, our black ring in Taub-NUT is given by the Maison data \eqref{finalchiandkappa}, with the parameters $\lambda$, $\gamma$, $b_4^{(2)}$ and $\alpha$ fixed as in equations \eqref{lambdasol}, \eqref{bsol}, \eqref{gammasolbis} and \eqref{alphasol} respectively. This completes our construction of the black ring in Taub-NUT. 

At this point it is worth summarizing the main steps in our construction:
\begin{enumerate}
\item Start form the seed metric (\ref{eqn:metricxy}) and construct the corresponding Maison data $\chi$ and $\kappa$, as shown in the appendix \ref{sec:appMaison}.
\item Apply the $SL(3,\mathbb{R})$ transformation (\ref{newchiandkappa}).
\item Fix the parameters $\lambda$, $b_4$ and $\gamma$ to the values given in (\ref{lambdasol}), (\ref{bsol}) and (\ref{gammasolbis}).
\item Apply the transformation (\ref{finalchiandkappa}), with $\alpha$ given in (\ref{alphasol});
\item Reconstruct the metric from $\chi''$ and $\kappa''$.
\end{enumerate}

The final solution can be written in the form:
\begin{equation}
\label{eqn:finalsol}
\begin{aligned}
ds^2&=\lambda'_{11}\big(d\xi^1+\mathbf A\big)^2-\frac{\tau'}{\lambda'_{11}}\big(dt+\mathbf B\big)^2\\
&\hspace{0.5cm}+\frac{1}{\tau'}\,\frac{\ell^2\hat R^4}{(x-y)^4}
	\bigg\{\frac{H(x,y)}{(1+b_4^2)H(y,x)}\,\big[F(x,y)-F(y,x)+2\,J(x,y)\big]
	\bigg(-\frac{dy^2}{G(y)}+\frac{dx^2}{G(x)}\bigg)\\
&\hspace{3.5cm}-G(x)G(y)\,d\phi_-^2\bigg\}\;,
\end{aligned}
\end{equation}
where
\begin{equation}
\mathbf A=\frac{\lambda'_{01}}{\lambda'_{11}}\,dt+2\,\frac{\lambda'_{01}\,\omega'^0_{-}+\lambda'_{11}\,\omega'^1_{-}}{\lambda'_{11}}\,d\phi_-\;,\qquad
\mathbf B=2\,\omega'^{0}_-\,d\phi_-\;, \label{eqn:final1forms}
\end{equation}
and the primed functions are obtained after the sequence of transformations described in the previous section. These can be read off from \eqref{finalchiandkappa}, which is given in terms of the $\chi$-matrix of the seed solution \eqref{eqn:metricxy} (see appendix \ref{sec:appMaison}). In \eqref{eqn:finalsol} and \eqref{eqn:final1forms} we have rescaled the $\phi_-$ angle, $\phi_-\to 2\,\phi_-$, so that $\Delta\phi_-=2\pi$.\footnote{The reason for doing this is that upon KK reduction along $\xi^1$, the angular coordinate $\phi_-$ becomes the four-dimensional azimuthal direction.} To obtain a regular (five-dimensional) solution, the parameters in the metric \eqref{eqn:finalsol} should obey
\begin{equation}
0< \nu < 1\;,\qquad 0< \hat R < \sqrt{2}\;. \label{eqn:ranges}
\end{equation}

\section{Exact extremal ring in Taub-NUT}
\label{sec:exactextr}

An appropriate extremal limit of the solution found in the previous
section gives an exact geometry representing a D0-D6 system at zero
temperature, with rotation only along the direction of the ring. The
solution has a singular horizon of vanishing area. By studying this
geometry for generic values of the distance between D0 and D6 charges,
we compute their exact interaction potential.

\subsection{The solution}

In the parametrization of section \ref{constructionsec}, the extremal solution is obtained by 
taking $\nu=0$: in this limit the horizon degenerates to a singular
point. The solution one obtains has only one independent angular momentum, corresponding
to rotation along the ring direction. 
As in the case of black rings in flat space, this
$S^1$-rotating extremal solutions cannot be balanced: when $\nu=0$ the
condition of absence of conical singularities, eq. (\ref{lambdasol}),
has only the trivial solution $\lambda=0$. Hence the space-time 
has a conical singularity, localized along the rod
$\rho=0,\,z\in[0,R^2/2]$ (or $x=1$). 

When $\nu=0$, the construction of section  \ref{constructionsec} drastically simplifies.
In particular, eq. (\ref{beq}), which guarantees that the $t$
components of rods 1 and 3 be aligned, is satisfied for any value of $b_4$. Thus the parameter $b_4$ remains unfixed, and one
has, in principle, a valid extremal black ring for any value of $b_4$.
It turns out, however, that solutions with different values of $b_4$ are
just different parametrizations of the same physical solution. The
solution with $b_4=0$ gives the simplest parametrization, and it is the
one that we will consider in the following. In this case, the value of
$\gamma$, given by  eq. (\ref{gammasol}), simplifies to
\be
\tan\gamma=\frac{\hat{R}}{2\sqrt 2}\,\sqrt{\frac{1+\lambda}{1-\lambda}}\,.
\ee
$\lambda$ remains a free
parameter of the solution, essentially determining its angular momentum.
The boost paremeter $\alpha$, needed to cancel the Dirac-Misner string, is fixed
to be
\be
\tanh\alpha=\hat R\,\frac{4(1-\lambda)+\lambda\hat R^2}{4(1-\lambda)-\lambda\hat R^2}\,
	\sqrt{\frac{1+\lambda}{8(1-\lambda)+\hat R^2(1+\lambda)}}\;.
\ee

Reconstructing the metric from the Maison data $\chi''$ and $\kappa''$,
and performing some trivial rescaling of coordinates, one can write the
metric of the extremal black ring in Taub-NUT in the form
\be
ds^2=g_{11}(d\xi^1 + A^1_0 dt + A^1_- d\phi_-)^2+ g_{11}^{-1/2}\Bigl[ -V (dt + A^0_- d\phi_-)^2+ V^{-1}
(e^{2 K} ds^2_B + {\hat \rho}^2 d\phi_-^2) \Bigr] \,.
\label{extremalBR}
\ee
The part of the metric in square brackets is the 4D Einstein-frame
metric. The metric depends on the parameters $\ell$ and $R$, which have
dimensions of length, and on the dimensionless parameter $\lambda$.  As before, we denote $\hat R=R/\ell$.

To write the metric coefficients
in compact form, it is convenient to define the dimensionless constants
\be
\mathcal{C}_1=\big[4(1-\lambda)+\lambda\hat{R}^2 \big]^2 \,,\quad \mathcal{C}_2 = \big[4(1-\lambda)+\hat{R}^2\big]^2 ,
\ee
and  the functions
\begin{equation}
\begin{aligned}
F_0=&~x+y+\lambda(1+xy)\, \\
F_1=&~\lambda \mathcal{C}_2 \big[(1-\lambda)(1-x)^2-(1+\lambda)(1+y)^2 \big]+\mathcal{C}_1 (1+\lambda) F_0\, \\
F_2=&~\mathcal{C}_2 F_0-\mathcal{C}_1\big[\lambda(1+x)^2+2(1-\lambda)x\big]\,  \\
F_3=&~\lambda \mathcal{C}_2\big[(1-\lambda)\big(x^2(x+y)-2x\big)+(1+\lambda)\big(y^2(x-y)+2(1-y^2)\big)
	+2\lambda(-2+x-y+2xy)\big]  \\
&- \mathcal{C}_1 (1+\lambda) \big[\lambda(x-y)(1+ xy)-(1-2\lambda)x^2-y^2+2(1-\lambda)\big] \,.
\end{aligned}
\end{equation}
Then the metric coefficients are given by
\bea
g_{11}&=&{(\mathcal{C}_2-\mathcal{C}_1) F_1\over \big[(1+\lambda) \mathcal{C}_1 -2\lambda  \mathcal{C}_2\big] F_2}\,,\nonumber\\
V&=&-\sqrt{\big[(1+\lambda) \mathcal{C}_1 - 2\lambda \mathcal{C}_2\big](\mathcal{C}_2-\mathcal{C}_1)} \,{F_0\over (F_1 F_2)^{1/2}}\,,
\nonumber\\
A^1_0&=&\lambda\sqrt{\frac{\mathcal{C}_2(1-\lambda^2)}{\mathcal{C}_2-\mathcal{C}_1}}\, \frac{(x-y)\big[\mathcal{C}_1 (1-x)+\mathcal{C}_2 (x+y)\big]}{F_1}\,,\nonumber\\
A^1_-&=&\frac{\ell\, \hat R^2}{2(\mathcal{C}_2-\mathcal{C}_1)}\sqrt{\frac{\mathcal{C}_1[(1+\lambda) \mathcal{C}_1 -2\lambda  \mathcal{C}_2]}{1-\lambda}}\,\frac{F_3}{(x-y)F_1}\,,\nonumber\\
A^0_-&=&\ell\, \hat{R}^2 \sqrt{(1+\lambda) \mathcal{C}_1 \mathcal{C}_2\over [(1+\lambda) \mathcal{C}_1 -2\lambda  \mathcal{C}_2] (\mathcal{C}_2 - \mathcal{C}_1)} \,{\lambda (1-x^2)(1+y)\over  (x-y) F_0}\,,\nonumber\\
e^{2 K}&=& {F_0\over(1-\lambda)( x+y)}\,,\nonumber\\
{\hat \rho}^2&=&  \ell^2\,\hat R^4\,{(1-x^2)(y^2-1)\over  (x-y)^4}\,,\nonumber\\
ds^2_B&=& -\ell^2\,\hat R^4\,{(x+y)\over (x-y)^3}\left({dx^2\over 1-x^2}+{dy^2\over y^2-1}\right)\,.
\eea

\subsection{Physical parameters}
The mass, charges and angular momentum of the solution are
\bea
M_\mathrm{tot}&\!\!=\!\!&\frac{\ell\, {\hat R}^2}{8\,G_4}\, {\mathcal{C}_1^2(1-\lambda) + 2\lambda (\mathcal{C}_2-\mathcal{C}_1)^2\over (\mathcal{C}_2-\mathcal{C}_1)[(1+\lambda) \mathcal{C}_1 -2\lambda  \mathcal{C}_2] } \,,
\nonumber\\
P&\!\!=\!\!&\frac{\ell\, \hat R^2}{4(\mathcal{C}_2-\mathcal{C}_1)}\,\sqrt{\frac{\mathcal{C}_1[(1+\lambda) \mathcal{C}_1 -2\lambda  \mathcal{C}_2] }{1-\lambda}}\,,\nonumber\\
Q&\!\!=\!\!&\frac{\ell\,\hat R^2\lambda}{2\big[(1+\lambda) \mathcal{C}_1 -2\lambda  \mathcal{C}_2\big]}
	\,\sqrt{\frac{(1+\lambda)\mathcal{C}_2(\mathcal{C}_2-\mathcal{C}_1)}{1-\lambda}}\,,
\nonumber\\
J&\!\!=\!\!&\frac{\ell^2\hat R^4}{8\,G_4}\,\frac{\lambda}{1-\lambda}
	\,\sqrt{\frac{(1+\lambda)\mathcal{C}_1\mathcal{C}_2}{(\mathcal{C}_2-\mathcal{C}_1)[(1+\lambda) \mathcal{C}_1 -2\lambda  \mathcal{C}_2]}}\,.
\label{charges}
\eea
($M_\mathrm{tot}$ is the total mass, and thus includes the magnetic monopole  contribution. This is different from $M$ in \eqref{genmass}.) Note that, as $\mathcal{C}_2>\mathcal{C}_1>0$ and $0\leq \lambda<1$, 
the combination $(1+\lambda) \mathcal{C}_1 -2\lambda  \mathcal{C}_2$ can
become negative for some value of the parameters. This happens for  $\hat{R}^2>\hat{R}_0^2$, where
\be
\hat{R}^2_0 =\frac{4(1-\lambda)}{2+\lambda}\left[-1+\sqrt{\frac{2(1+\lambda)}{\lambda}}\right]\,\,.
\ee
Beyond this point the magnetic charge becomes imaginary, and the metric ceases to make sense. 
Thus the parameters have to be taken in the range
\be
\ell>0\,,\quad 0\leq \hat R<\hat{R}_0 .
\label{range}
\ee
For this range of parameters the magnetic and electric charges $P$ and
$Q$ attain all values from zero to $+\infty$, and thus the range
(\ref{range}) covers the whole physical spectrum of the charges.

Note however that the solution \eqref{extremalBR} only spans a codimension one subspace of the full phase space of extremal black rings in Taub-NUT, describing rings  for which the angular momentum is linked to the charges as
\be
P Q =G_4 J \,.
\ee
The condition above restricts to
configurations that, when uplifted to 5D, have angular momentum in only
one plane, which turns out to be the plane of the ring.
 
\subsection{Limits}
\label{sec:limits}
In this subsection we analyze the various limits that connect the metric  \eqref{extremalBR} to previously known solutions. Taking the radius of the ring to zero one reproduces the extremal KK black hole with $G_4J=PQ$ found in \cite{DR1,FL1}. We also show how to recover the extremal
$S^1$--spinning ring in flat space found in \cite{ER1}. This limit
corresponds to localising the black ring near the tip of Taub-NUT space
and zooming into that region.

\subsubsection*{Extremal KK black hole with $G_4 J=PQ$}
We can recover the extremal KK black hole of \cite{DR1,FL1} with $G_4
J=PQ$ as the $R\to 0$ limit of our extremal black ring in Taub-NUT. One should keep in mind 
that the KK black hole with $G_4J=PQ$ has zero
area and, strictly speaking, should be regarded as a naked singularity.

Recall that from five-dimensional viewpoint, KK black holes with non-zero magnetic charge can be thought of black holes sitting at the tip of the Taub-NUT space. Therefore,  this limit of our solution can be regarded as a zero-radius limit in which effectively we are localizing the black ring at the tip of the Taub-NUT space while keeping the radius of the Taub-NUT circle, and hence the magnetic charge $P$, as well as the electric charge $Q$, fixed.  This is achieved by taking $R\to 0$ and $\lambda\to 1$ keeping fixed the parameters $\tilde R$ and $\ell$, with
\begin{equation}
\tilde R=\frac{R}{\sqrt{1-\lambda}}\,.
\end{equation}
One should also change coordinates $(x,y)\to (r,\theta)$ as
\begin{equation}
x\to -1+\frac{R^2}{\ell r}\,\cos^2\left(\textstyle\frac{\theta}{2}\displaystyle\right)\,,~~~~~~
y\to -1-\frac{R^2}{\ell r}\,\sin^2\left(\textstyle\frac{\theta}{2}\right)\,.
\end{equation}
The resulting metric is
\begin{equation}
 ds^2=\frac{H_q}{H_p}\left(d\xi^1+\mathbf A\right)^2-\frac{r^2}{H_q}\left(dt+\mathbf B\right)^2
	+H_p\left(\frac{dr^2}{r^2}+d\theta^2+\sin^2\theta\,d\phi^2\right)\,,
	\label{eqn:kkbh}
\end{equation}
where
\begin{subequations}
 \begin{align}
H_p&=r^2+r\,p+\frac{p^2 q}{2(p+q)}\,(1+\cos\theta)\,,~~~~~~
H_q=r^2+r\,q+\frac{p\,q^2}{2(p+q)}\,(1-\cos\theta)\,,\\
\mathbf A&=-\frac{1}{H_q}\bigg\{Q\big[2r+p(1-\cos\theta)\big]dt+P\Big[2H_q\cos\theta-q\Big(\textstyle r+\frac{p\,q}{p+q}\displaystyle\Big)\sin^2\theta\Big]d\phi\bigg\}\,,\\
\mathbf B&=\frac{(p\,q)^{3/2}}{2(p+q)r^2}\,\sin^2\theta \,d\phi\,.
 \end{align}
\end{subequations}
The parameters $p$ and $q$ (with $p>0$, $q>0$) are related to $\tilde R$ and $\ell$ as
\be
\tilde R=\frac{p\sqrt{q(4p+3q)}}{2(p+q)}
\,,~~~~~~\ell=\frac{p(4p+3q)}{4(p+q)}\,,
\label{qpvar}
\ee
 and the electric and magnetic charges are given by
\begin{equation}
Q^2=\frac{q^3}{4(p+q)}\,,~~~~~~P^2=\frac{p^3}{4(p+q)}\,.
\label{rasheedcharges}
\end{equation}
The metric \eqref{eqn:kkbh} is just the $a=m=0$ KK black hole in the form presented in \cite{FL1}.  Moreover, one can check that in this limit the charges satisfy
\begin{equation}
2 G_4 M_\mathrm{tot}=\big[Q^{2/3}+P^{2/3}\big]^{3/2}\,.
\label{rasheedmass}
\end{equation}

\subsubsection*{Extremal $S^1$--spinning black ring in flat space}
In the limit in which the KK radius becomes much larger than the ring
size, one expects to recover the extremal $S^1$--spinning black ring in
flat space . This limit is achieved by sending the magnetic charge $P$
to infinity while keeping finite the size of the black ring, which is
roughly given by $R$. In our parametrization, we have to send
$\ell\to\infty$ while $R$, $\lambda$ and the coordinates $(x,y)$ are
kept fixed. To recover the extremal limit of the $S^1$--spinning ring of
\cite{ER1} in the form presented in \cite{RE1}, one also has to redefine
the $R$ parameter as $R\to R\sqrt{\frac{1-\lambda}{2}}$ and change the
angular coordinates as
\begin{equation}
\phi_+=\frac{1}{\sqrt{1-\lambda}}(\phi'-\psi')\,,~~~~~~\phi_-=
\frac{1}{\sqrt{1-\lambda}}(\phi'+\psi')\,,
\end{equation}
since the angular coordinates $(\phi',\psi')$ of \cite{RE1} are not
canonically normalized. Then we obtain:
\begin{equation}
\begin{aligned}
ds^2=&-\frac{F(y)}{F(x)}\left(dt-C\,R\,\frac{1+y}{F(y)}\,d\psi'\right)^2\\
&+\frac{R^2}{(x-y)^2}\,F(x)\left[-\frac{G(y)}{F(y)}\,d{\psi'}^2-\frac{dy^2}{G(y)}+\frac{dx^2}{G(x)}+\frac{G(x)}{F(x)}\,d{\phi'}^2\right]\,,
\end{aligned}
\end{equation}
where $C=\lambda\sqrt{\frac{1+\lambda}{1-\lambda}}$, $F(\xi)=1+\lambda
\xi$ and $G(\xi)=1-\xi^2$. This is the $\nu\to 0$ limit of the metric
given in \cite{RE1}.

\subsection{Interaction energy}
A measure of the interaction energy between the D0 and D6 charges is given by 
\be
E_\text{int}=M_\mathrm{tot}-(M_\text{D0}+M_\text{D6})= M_\mathrm{tot}-{Q+P\over 2 G_4}\,,
\ee
where $M_\mathrm{tot}$ is the total mass of the system, measured at
infinity as the ADM mass, and $M_\text{D0}$,
$M_\text{D6}$ are the masses 
of the D0 and D6 branes in isolation. 

To obtain a physical understanding of $E_\text{int}$ one should express
it in terms of the charges $Q$ and $P$ and of the distance between them.
A rough estimate of this distance is given by the parameter $R$. A more
precise measure of distance, at least in the limit of large separation,
when the distortion on the metric due to the interaction between D0 and
D6 is small, is furnished by the length of the rod at $\hat{\rho}=0$ and
$0<\hat{z}<R^2/(2 \ell)$, computed with the 5D metric
(\ref{extremalBR}):
\begin{equation}
R_{\text{ph}}=\int_{-\infty}^{-1}dy\sqrt{\left.
G^{(5)}_{yy}\right|_{x\rightarrow 1}} \;,
\end{equation}
where $G^{(5)}_{yy}$ is the $yy$ component of the 5D metric (\ref{extremalBR}): 
\begin{equation}
G^{(5)}_{yy}={R^4\over \ell^2} g_{11}^{-1/2} V^{-1} e^{2 K}
\frac{(x+y)}{(x-y)^3(1-y^2)}.
\end{equation}
In the limit $R\to0$, with fixed $Q$ and $P$, one finds that
$R_\text{ph}$ so defined goes to a non-zero value, given by
\be
R_\text{min}= 4 Q^{1/3} P^{2/3}\,.
\ee
This is a quite counterintuitive result: as we have shown in section
\ref{sec:limits}, the black ring reduces in this limit to the extremal
KK black hole, which, naively, represents the configuration in which the
D0 and D6 charges are on top of each other. The fact that one finds
instead a non-zero distance $R_\text{min}$ can be attributed to the
large distortion on the metric due to the D0-D6 interaction (we will
show that in this limit the metric has the maximal conical defect angle
$\Delta=2\pi$). To correct for this effect, we redefine
$R_\text{ph}$ as $R_\text{ph}\to R_\text{ph}-R_\text{min}$.

The behavior of $E_\text{int}$ as a function of $R$,
for different values of the ratio
$Q/P$, was shown in fig.~\ref{fig:D0D6noB}.
$E_\text{int}$ is a monotonically decreasing function of the distance,
that goes to a positive value for $R=0$ and vanishes for large $R$.
Hence the interaction between D0 and D6 branes is always repulsive. The
two limits of $R$ small and large $R$ can be understood
analytically.

The limit of small $R$ and fixed charges is the same as the limit in
which the black ring reduces to the KK black hole. As the mass of the KK
black hole is given by (\ref{rasheedmass}), the interaction energy in
this limit is given by
\be
2 G_4 E_\text{int}\approx (Q^{2/3}+P^{2/3})^{3/2}-(Q+P)>0\,.
\ee
Keeping the terms of higher order in $R$, one can see that
$E_\text{int}$ has an extremum at $R=0$ as a function of $R$. On the
other hand, when expressed in terms of $R_\text{ph}$, $E_\text{int}$ has
a non-vanishing slope at $R_\text{ph}=0$: this is due to the fact that
the physical distance $R_\text{ph}$ depends quadratically on $R$, for
small $R$. However, as we have already noted above, $R_\text{ph}$ does
not seem to provide a good measure of the distance between D0 and D6 in
this limit.

The opposite limit is when $R\gg Q,\,P$. To achieve this limit one
should take $R/\ell$ large; note however that this cannot be done at
fixed $\lambda$, due to the constraint (\ref{range}). One should also
adjust $\lambda$ in such a way that the upper bound for $R/\ell$, $\hat
R_0$, becomes large, which happens if $\lambda$ goes to zero. Thus the
appropriate limit is
\be
R = {\ell\over \sqrt{\epsilon}}\,,\quad \lambda=\lambda_0 \epsilon^4\,,\quad \epsilon\to 0\,.
\label{larger}
\ee
In this limit one has
\be
{Q\over P}\approx {\lambda_0\over 128}\,,\quad P \approx 4\epsilon \ell\,,\quad G_4 M_\mathrm{tot}\approx 2 \ell \Bigl(1+{\lambda_0\over 128}\Bigr)\epsilon+{\ell\over 4}\lambda_0 \epsilon^3\,.
\label{largerbis}
\ee
These relations imply that the interaction potential is given by
\be
G_4 E_\text{int} =  2^{1/3} Q \Bigl({P\over R}\Bigr)^{4/3}\,.
\ee
We can re-express this result in terms of the physical distance $R_\text{ph}$, which, for large separations,\footnote{At large distances the five-dimensional physical distance coincides with the four-dimensional one (computed in Einstein frame) to leading order.}  is given by
\be
R_{\text{ph}}\approx 2^{-1/3} P \Bigl({R\over P}\Bigr)^{4/3}\,.
\label{largerRph}
\ee
Substituting this $R_{\text{ph}}$ in the above expression for $E_\text{int}$, one  
finds 
\be
G_4 E_\text{int} =  {Q P\over R_\text{ph}}\,,
\ee
which has the form of a repulsive Coulomb potential between charges $Q$
and $P$. This result is in agreement with the interaction energy
(\ref{eintb=0}) derived by the perturbative method.

\subsection{Conical defect}
Further information on the interaction between the D0 and D6 charges are obtained by analyzing the conical defect singularity of the extremal solution (\ref{extremalBR}). This singularity is what is needed to counter-balance the repulsion between the D0's and the D6's, and thus the stress tensor associated to it gives a measure of the interaction.

Following a method used in \cite{CP} in an analogous context, we will compute the delta-function-like
contribution to the Ricci tensor at the conical singularity, and use Einstein's equation to derive the associated stress tensor. 

To obtain a duality invariant description of the system, we focus on the 4D Einstein frame metric, given by
\be
ds^2_4= - V  (dt +A^0 d\phi_-)^2 + V^{-1}\Bigl[ e^{2 K} (d\hat{\rho}^2 + d{\hat z}^2) +{\hat \rho}^2 d\phi_-^2\Bigr]\,,
\ee
where $\hat{\rho}$ and $\hat{z}$ are the 4D Weyl coordinates
\be
\hat{\rho}={\rho\over \ell}\,,\quad \hat{z} = {z\over \ell}\,.
\ee
In the vicinity of the rod  $\hat{\rho}=0$ and $0<\hat{z}<R^2/(2 \ell)$, one has that
\be
A^0\approx 0\,,\quad e^{2 K} \approx {1+\lambda\over 1-\lambda}\,.
\ee
Thus near this rod the metric in the $\hat{\rho}$-$\phi_-$ plane decouples from the remaining directions and becomes conformally equivalent to the flat metric in $\mathbb{R}^2$ after the change of coordinates
\be
\hat{\phi}={\phi_-\over e^{K}}= \sqrt{1-\lambda\over 1+\lambda} \phi_-\,.
\ee 
The condition that at asymptotic infinity the metric be flat fixes the periodicity of $\phi_-$ to be $2\pi$. We thus see that along the rod under consideration there is a conical defect given by 
\be
\Delta= 2\pi \Bigl(1- {1\over  e^{K}}\Bigr) = 2\pi \Bigl(1-\sqrt{1-\lambda\over 1+\lambda}\Bigr)\,.
\ee
Note that for our geometries the deficit angle $\Delta$ is always positive.

The curvature due to this conical defect can be computed from the general relation \cite{TR1}
\be
\int_\mathcal{M}  R = 2 \Delta \mathcal{A}_\Sigma \,,
\label{regge}
\ee 
where $R$ is the Ricci scalar, $\cal{M}$ is the full space-time manifold, $\mathcal{A}_\Sigma$ is the space-time area of the conical defect. In our case $\Sigma$ is the rod $0<\hat{z}<R^2/(2 \ell)$ times time, and its area is $\mathcal{A}_\Sigma = \int   dt d\hat{z}\,e^{K}$.  Thus, eq. (\ref{regge}) implies
\be
R = 2 \Delta V \delta(\hat{\rho}) \,,
\ee
where the delta-function $\delta(\hat{\rho})$ is normalized as $\int d\hat{\rho} d\phi_- \,{\hat \rho}\,e^K \delta(\hat{r})   =1 $. From the Ricci scalar $R$, and the fact that the curvature only has components along the $\hat{\rho}$-$\phi_-$ plane  --- so that $R_{00}=R_{zz}=0$ --- one can derive the Einstein tensor and hence the stress tensor $T_{\mu\nu}$ associated to the conical defect. One finds that
\be
T_{00} = (8\pi G_4)^{-1} \Delta V^2  \delta(\hat{\rho}) \,,\quad T_{zz} = -(8\pi G_4)^{-1}\Delta e^{2 K}  \delta(\hat{\rho}) \,.
\ee
The energy of the conical defect, obtained by integrating the energy density  ${T_{00}\over \sqrt{- g_{00}}}$ over the space directions, is given by
\be
E_\text{def}= \int d\hat{z} d\hat{r} d\phi_-\; {\hat\rho}\,e^{2 K}\,{T_{00}\over \sqrt{- g_{00}}}\,V^{-3/2}   =(8 G_4)^{-1} \bigg(\sqrt{1+\lambda\over 1-\lambda}-1\bigg) {R^2\over \ell} .
\label{eenergy}
\ee
The force exerted by the strut is obtained as the integral of the pressure ${T_{zz}\over g_{zz}}$ over the directions transverse to the strut:
\be
F_\text{def}=\int d\hat{\rho} d\phi_-\;{\hat\rho}\,e^K \,{T_{zz}\over g_{zz}}\, V^{-1}  =  -(4 G_4)^{-1}\bigg(1- \sqrt{1-\lambda\over 1+\lambda}\bigg) \,.
\label{epressure}
\ee
We note that $F_\text{def}$ is proportional to the conical defect $\Delta$.

\begin{figure}[t]
\centering
\includegraphics{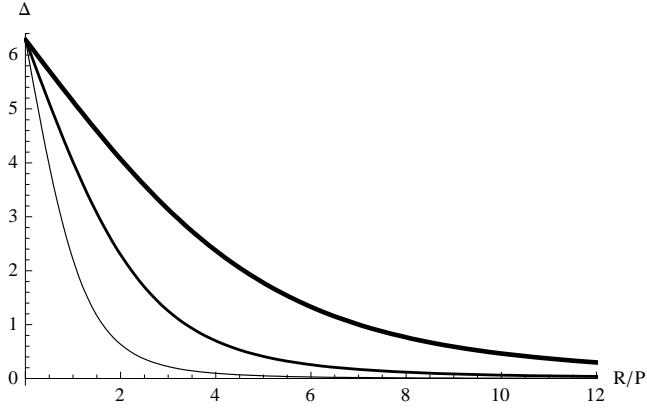}
\caption{\small{Conical defect versus radius $R$ in units of $P$ for $Q/P=.1$ (thin), $1$ (thick) and $10$ (thicker). Note that the conical defect is never zero for finite values of $R$. This means that the solution is never balanced. In the $R\rightarrow 0$ limit the conical defect is maximal: $2\pi$. The force $F_\text{def}$ is porportional to $\Delta$ (\ref{epressure}).}}
\label{pdeltavsR}
\end{figure}

\begin{figure}[h!]
\centering
\includegraphics{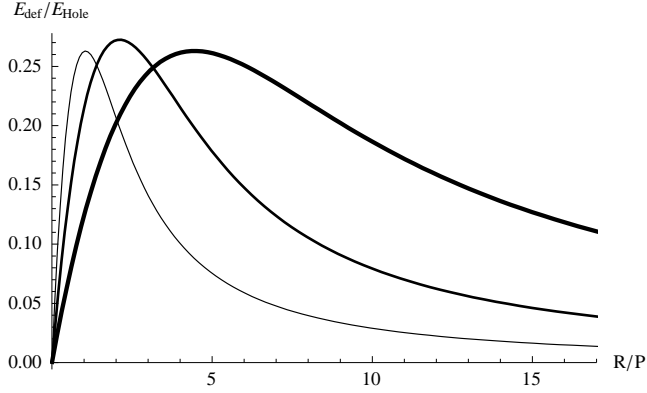}
\caption{\small{Energy of the conical defect versus $R/P$ for $Q/P=.1$ (thin),
$1$ (thick) and $10$ (thicker). Energies are normalised as in figure
\ref{fig:D0D6noB}. The plot shows that $E_\text{def}$ has a maximum at
some finite value of $R$. The position of the maximum increases with
$Q/P$. The vanishing of $E_\text{def}$ at small $R$ is a volume effect,
and does not mean that the force between the charges becomes attractive
at small distances.}}
\label{EvsR}
\end{figure}

The plots describing the behavior of $E_\text{def}$ and $F_\text{def}$ as a function of the distance parameter $R$, for fixed values of the charges $Q$ and $P$ are depicted in figures \ref{pdeltavsR} and  \ref{EvsR}. 

One notes from these plots that the force is always repulsive and is maximal at $R=0$. The energy $E_\text{def}$ vanishes at large $R$, has a maximum at some finite value of $R$ and vanishes again at $R=0$. This seems to contradict the behavior found for the interaction potential $E_\text{int}$, which gave a repulsive potential for every value of $R$. We interpret the vanishing of $E_\text{def}$ at small $R$ as a volume effect: at $R=0$ the metric has a maximal conical defect $\Delta=2\pi$, and thus the space transverse to the strut becomes effectively one-dimensional, and its volume vanishes. Hence the fact that $E_\text{def}$ vanishes does not mean that the D0-D6 interaction becomes attractive at small $R$.

The behavior of $E_\text{def}$ and $F_\text{def}$ in the limits of small and large separations can be reproduced analytically. For small $R$ one finds
\bea
8G_4E_\text{def} &\approx& 2 \sqrt{2}R  {Q^{1/3}\over\sqrt{4 P^{2/3}+3 Q^{2/3}}}\,,\nonumber\\
4G_4 F_\text{def} &\approx& 
-1+ {R\over P^{2/3} Q^{1/3}}\sqrt{P^{2/3}+Q^{2/3}\over 2  (4 P^{2/3}+3 Q^{2/3})}\,.
\eea

The large $R$ behavior is given by
\bea
8 G_4 E_\text{def}&\approx& 2^{7/3} Q \Bigl({P\over R}\Bigr)^{4/3}\,,\nonumber\\
4 G_4 F_\text{def}  &\approx& - 2^{5/3} {Q\over P}\Bigl({P\over R}\Bigr)^{8/3} .
\label{enplarger}
\eea
Once expressed in terms of the physical distance $R_\text{ph}$, given in this limit in (\ref{largerRph}), these expressions simplify to 
\bea
2 G_4 E_\text{def}&\approx& {Q P\over R_\text{ph}}\,,\nonumber\\
2 G_4 F_\text{def}  &\approx&  -{Q P\over R^2_\text{ph}}\,,
\eea
which are again of Coulombic form. Note that, in this limit, the energy of the conical defect accounts for half of  the interaction energy between D0 and D6 branes: $E_\text{def}=E_\text{int}/2$.

\section{Exact black rings in Taub-NUT}
\label{sec:exactrings}

Using the exact solution constructed in section \ref{constructionsec},
we can study the D0-D6 system above extremality in a regime in which the
gravitational backreaction of the D0 branes on the D6 Taub-NUT
background is fully taken into account. We are interested in
configurations of equilibrium, in which conical singularities have been
eliminated.

In this analysis, one should keep in mind that the solution of section
\ref{constructionsec} has only one independent angular momentum, and
thus it does not represent the most general black ring in Taub-NUT.
Indeed, the starting seed solution \eqref{eqn:metricxy} only has
intrinsic rotation along the $S^1$ of the ring. This can be seen from
the fact that only the BZ transformation affecting the horizon rod gives
rotation to the solution. The angular momentum on the $S^2$ is induced
by the soliton transformation on the spacelike rods, which mixes the
$\hat\phi$ and $\hat\psi$ directions. It is precisely this lack of generality
that implies that the four-dimensional conserved charges cannot all be
independent. Recall that from a four-dimensional perspective, our
solution is characterized by the four conserved charges
$(M_\mathrm{tot},\,J,\,Q,\,P)$, see appendix \ref{app:charges}. However, our
(balanced) solution only has three free parameters, namely, an overall
length scale $\ell$ and two dimensionless quantites, $(\hat R,\nu)$.
Therefore, once we have fixed this overall scale fixing, say, $P$, there
must exist a relation between the remaining conserved charges, which
relates the angular momentum $J$ along the $S^2$ to $Q$, $P$ and
$M_\mathrm{tot}$. At the limiting endpoints of the family of solutions,
\ie when the ring
is infinitely far from the nut, or
when it collapses into a singular extremal black hole,
this relation becomes $J= P Q/G_4$, but at any other point in the space
of our solutions we have $J\neq P Q/G_4$, signaling a non-zero component
along the $S^2$ of the ring.

A more general doubly spinning black ring in Taub-NUT must exist with
independent rotation along the $S^2$ which should allow to vary $J$
independently of $PQ$, so the four dimensional solution would be
characterized by four independent conserved charges. In fact such
solutions are easily described in the thin ring limit within the
perturbative approach, see appendix~\ref{app:doublespin}. In this paper we
shall content ourselves with studying in detail the physics of the black
ring in Taub-NUT with just one independent angular momentum. We leave
the problem of constructing the general solution for future work. 

From the four-dimensional perspective, we have an electrically charged
rotating black hole at $y=-\frac{1}{\nu}$, separated from the magnetic
monopole, which sits at $(x=+1,y=-1)$ and which appears as a naked
singularity, although the five-dimensional geometry is regular.

\subsection{Dimensionless quantities} 

We characterize the solution in terms of four-dimensional magnitudes,
since the solution is asymptotically flat in the non-compact four
dimensions.

In Kaluza-Klein solutions it is natural to fix the
length of the KK circle and measure all dimensionful quantities relative
to it. Since the asymptotic
length of this circle is measured by the magnetic charge as in
\eqref{PQN}, an essentially equivalent way of doing this is to
define dimensionless quantities by dividing
the physical magnitudes by suitable powers of $P$. We first fix one of
the dimensionless parameters by fixing the total mass
 \begin{equation}
 \frac{G_4M_\mathrm{tot}}{P}\equiv \mu+\frac{1}{2}\;.
 \end{equation}
Note that $\mu\geq 0$, with $\mu=0$ when the mass equals
the D6 brane mass
 \begin{equation}
 M_\textrm{D6}=\frac{P}{2\,G_4}\;. \label{eqn:KKmass}
 \end{equation} 
Then $\mu$ measures the energy above the D6 brane mass. This is
convenient, since we regard the D6 brane as remaining unexcited, while
the D0 brane (the black ring) is thermally excited.

We define other dimensionless conserved quantities as
\begin{equation}
 a_\textrm{H}=\frac{\mathcal{A}_4}{P^2}\;,\qquad
 j=\frac{G_4\,J}{P^2}\;,\qquad  q=\frac{Q}{P}\;.
\end{equation}

If we are interested in having a five-dimensional perspective on the
solution,  we note, using the relations (\ref{jrelation}), that
the angular momenta $J_\psi$ along the $S^1$ and $J_\phi$ along the $S^2$ of the black ring
are captured by the five-dimensional dimensionless quantities
\begin{equation}
j_\psi=q\;,\qquad j_\phi=j-q\;,
\end{equation}
see eq.~\eqref{qjpsi}.
Thus $q$
measures the $S^1$ spin of the ring. The four and five-dimensional
horizon areas differ only in a constant factor since we are keeping $L$
fixed and so $a_H$ represents both.

\subsection{Phase diagram}

\begin{figure}[t!]
\begin{center}
\includegraphics[scale=0.65]{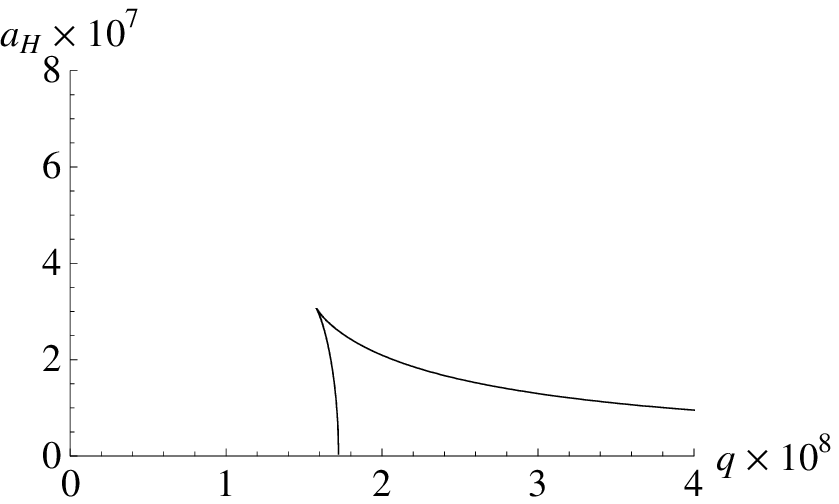}
\includegraphics[scale=0.65]{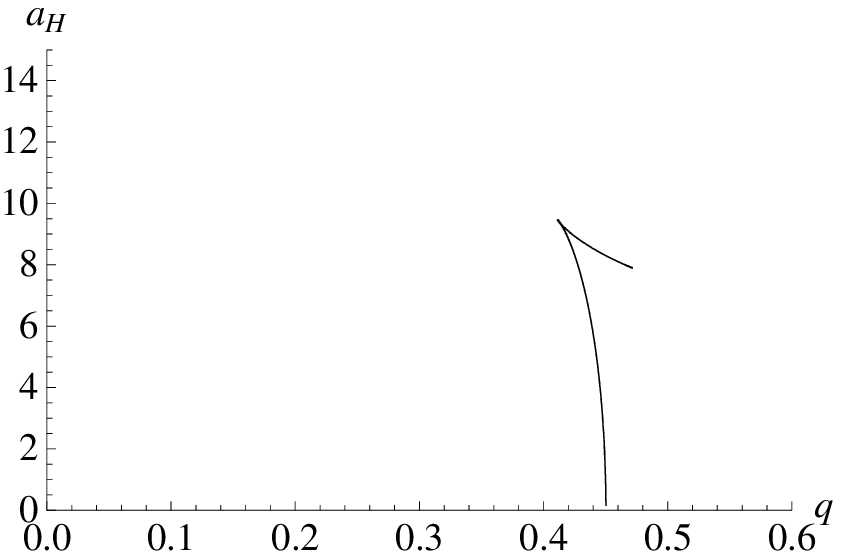}
\includegraphics[scale=0.65]{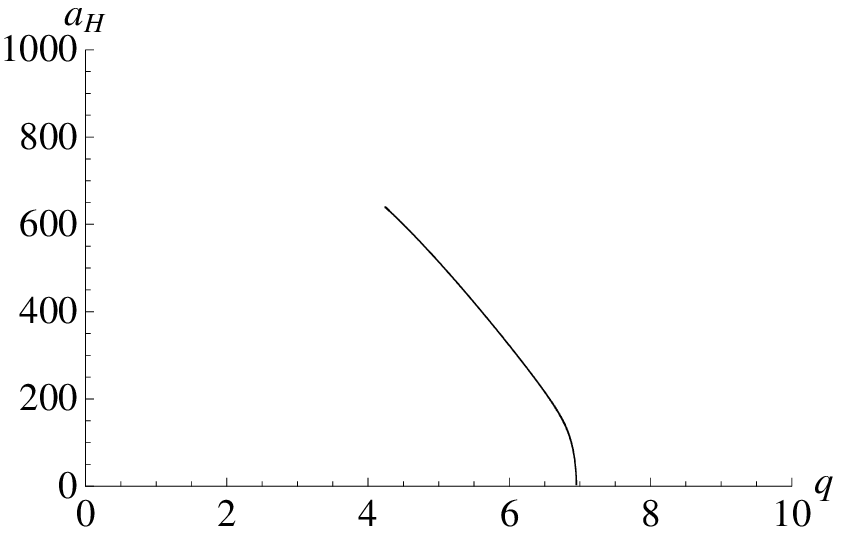}
\end{center}
\caption{\small{Phase diagram of black rings in Taub-NUT. \textit{Left:}
Phase diagram for $\mu=5\times 10^{-6}$. In this plot the value of the
reduced area $a_\textrm{H}$ and that of the reduced electric charge $q$
have been multiplied by a factor of $10^7$ and $10^8$ respectively.
\textit{Center:} Phase diagram for $\mu=1/2$. For small $\mu$ the curves
resemble the ones in asymptotically flat five dimensions, but in
contrast to them, the thin ring branch extends only up to a maximum
value of the charge for given mass. The limiting values of the area and
charge for fixed mass can be found from \eqref{acrit} and \eqref{Mcrit}.
\textit{Right:} Phase diagram for $\mu=9/2$. For $\mu>\mu_c=9/10$ the
thin ring branch disappears: we regard this as nothing more than a
peculiar feature of the particular family of solutions we have
constructed.}}
\label{fig:phasediagb42}
\end{figure}

For fixed magnetic charge $P$, the area of the black ring is a
two-dimensional surface over the plane of $\mu$ and $q$,
$a_\textrm{H}(\mu, q)$. In order to visualize it, we consider sections
at constant $\mu$. In figure
\ref{fig:phasediagb42} we present three illustrative plots, one at a very
small value of $\mu$, another at a value $\mu=1/2$ that makes the D0
and D6 equally heavy, and a third
one at very large D0 mass, $\mu=9/2$.

\begin{figure}[h]
 \begin{center}
  \includegraphics[scale=0.65]{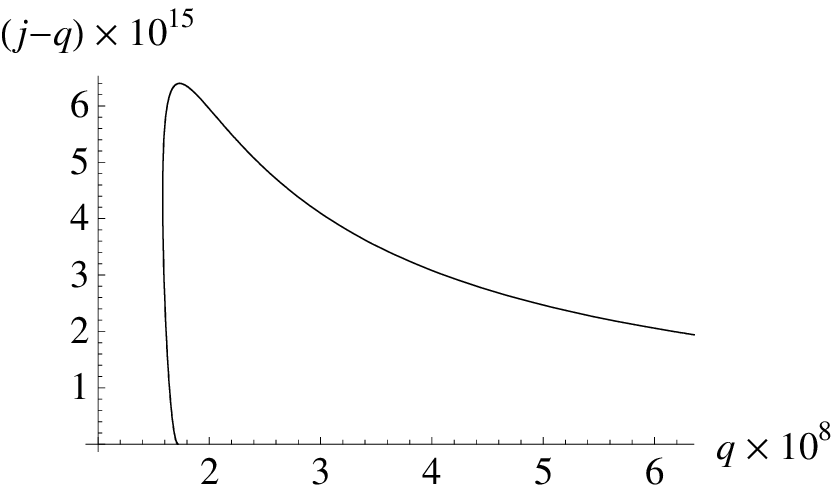}
\includegraphics[scale=0.65]{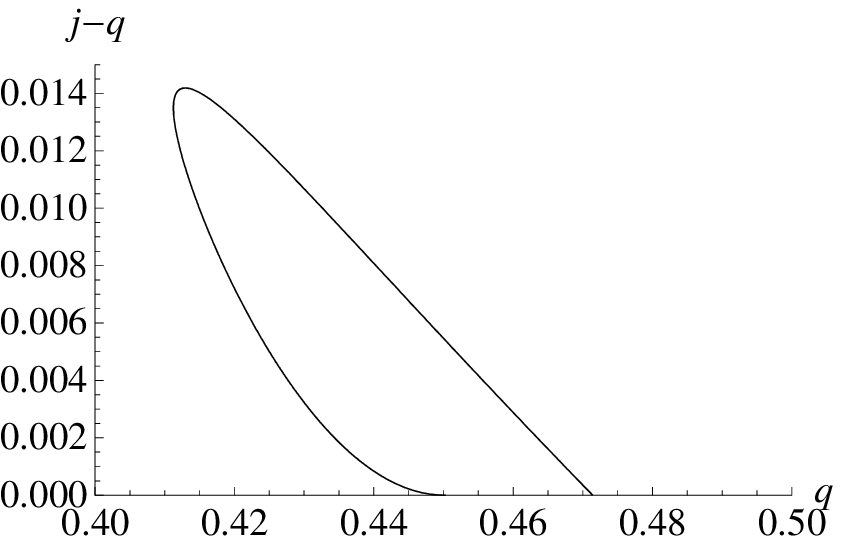}
\includegraphics[scale=0.65]{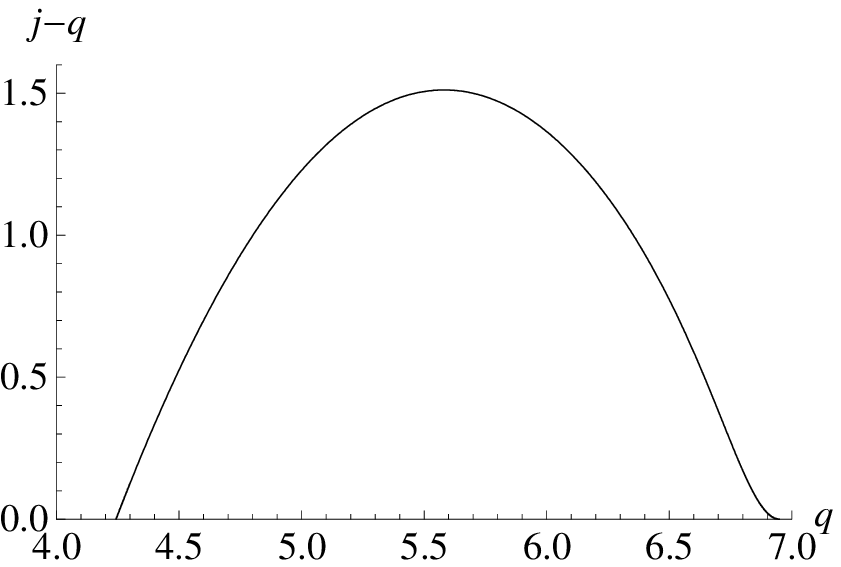}
 \end{center}
\caption{\small{Phases in the $(q, j-q)$-plane. From left to right we
show the phases corresponding to $\mu=5\times 10^{-6}$, $1/2$ and
$\mu=9/2$ respectively. These curves represent only the subfamily
constructed in this paper of a larger family of solutions with
four-independent parameters, which should cover a finite region of the
$(q, j-q)$-plane at every $\mu$.}}
\label{fig:phasesj2vsj1}
\end{figure}

Just like in the asymptotically flat case, black rings exist in two
branches, usually referred to as thin and fat rings. Despite the
similarity to the singly spinning ring in flat space \cite{ER1}, there
is a small distorsion due to the fact that our black ring is doubly
spinning, and the angular momentum in the $S^2$ of the ring varies along
the curves shown in figure \ref{fig:phasesj2vsj1}. Therefore it is more
accurate to compare it to a family of doubly spinning black rings, whose
$S^2$-spin is generically small and vanishes at the
endpoint solutions.

For $\mu$ close to $0$ the black ring can be thought of as a small
perturbation in the KK monopole background. In this regime the phase
diagram looks very similar to that of a single-spin black ring in flat
space---observe in fig.~\ref{fig:phasesj2vsj1} that for small $\mu$ we
have $j\simeq q$ so $j_\phi\simeq 0$. However, the curves of $a_H$ at
fixed $\mu$ terminate at a finite value of the charge and area. The
endpoints can in fact be precisely calculated: since they correspond to
rings at infinite distance from the nut, they are accurately described
by the construction of sec.~\ref{sec:pertd0d6}. Thus the limiting values
of the area and charge for fixed mass correspond to the critical values
computed in \eqref{acrit} and \eqref{Mcrit}. This is the case in fact
for all values of $\mu$, not just small ones. For small $\mu$ we can also find the approximate values of the lower bound on $q$ and upper bound on $a_H$ using the bounds on black rings in flat space,
\beq
a_H\leq \frac{32\sqrt{2}\pi}{3\sqrt{3}}\mu^{3/2}\,,\qquad
q\geq \sqrt{2}\,\mu^{3/2}\qquad (\mu\ll 1)\,.
\eeq
These correspond in fact to the bounds quoted in \eqref{upperbounds}.

As $\mu$ increases the lower limit on the area for fixed mass increases
and the upper limit on the charge decreases (see \eqref{thinMAQ}), so
the thin ring branch shortens. As we move slightly away from this point
the angular momentum along the $S^2$ of the ring switches on. The
behavior of the phase diagram near the endpoint of the thin ring branch
for arbitrary values of $\mu$ can be analytically derived from our exact
solution: it is described by the limit in which the parameter $\nu$ is
sent to zero, and $\hat{R}$ is taken of the form $\hat{R}=\sqrt{2}(1-
\eta \nu)$, where $\eta$ is a fixed parameter controlling the mass of
the solution. In this limit the perturbative approximation of
sec.~\ref{sec:pertd0d6} becomes accurate, and indeed it is possible to
match the limit of our exact solution with a boosted Kerr string with
fixed boost parameter and perturbative angular momentum (see appendix
\ref{app:doublespin}). 

As we keep increasing $\mu$ we observe that the thin ring branch
disappears at the critical value of $\mu_c=9/10$. We believe that this is
an artifact of our (particular) solution and that a more general black
ring in Taub-NUT should have two branches of solutions for all
$\mu>0$, in agreement with the perturbative construction of section
\ref{sec:pertd0d6}, which clearly contains thin black rings far from the
nut for every value of $\mu$.

For all values of $\mu$, there is a particular limiting value of $q$ for
which $a_\textrm{H}\to 0$. This corresponds to the zero radius limit of
our solutions, and we recover the extremal $(G_4 J=PQ)$ nakedly
singular KK black hole of \cite{DR1,FL1}.

\section{Outlook}
\label{sec:outlook}

Our study provides an example of how novel gravitational techniques can
be applied to extract useful information about the dynamics of D-branes. In
particular, we have uncovered a new way ---via thermal excitation--- to
produce bound states of D0 and D6 branes, and exhibited how a certain
extremal black hole can be formed by bringing together D0 and D6 branes
in a continuous manner, with the black hole mass corresponding to the mass
of the D0 and D6 constituents plus the interaction energy.

We have also used our methods to analyze the stability of the
configurations. The reader may have noticed that we have only discussed
stability with respect to changes in the distance between the D0 and D6
branes, which is to say, changes in the radial position of the black
ring in the Taub-NUT background. Thin black rings are known to be stable
to such changes in asymptotically flat space \cite{EEV1}, but on the
other hand, they are expected to be generically {\it unstable} to
Gregory-Laflamme-type of modes that create inhomogeneities along the
ring and that would be missed by our analysis. Should not we expect our
black rings to suffer from them, too? 

The answer is that such instabilities can in fact be avoided. Extremal singular
rings, corresponding to zero-temperature D0 branes, are certainly not
expected to suffer from them. Non-extremal black rings should suffer
from GL instabilities only if they are thinner than the Kaluza-Klein
radius, $r_0<L$. However, we have seen that within the scope of our
methods we can study black rings with $r_0>L$ as long as they are far
from the nut, $R\gg r_0$. Such black rings resemble black strings that
are not afflicted by the GL instability. We have found that they are also
radially stable so, since no other mechanism for instability is known to
affect them, we can expect these black rings to be stable. While we have not
discussed in any detail {\it fat} black rings (with, roughly, $r_0\sim
R$), for example those that correspond to the lower branches of the
curves in fig.~\ref{fig:phasediagb42}, the generic arguments of
\cite{EEV1} lead us to expect them to be radially unstable.

There is a number of possible extensions of our work. For instance, the
perturbative techniques can be easily extended to other backgrounds of
Taub-NUT type, supersymmetric or otherwise, and also to charged black
rings (see \eg \cite{BEI1}). 

Perhaps more interestingly, with the present techniques for generating
exact solutions we could also obtain solutions where a black hole sits
at the nut. This would correspond to thermally exciting the D6 branes,
and presumably it would allow for equilibrium states of extremal or
non-extremal D0 branes a finite distance apart. In five dimensions, the
configuration can be regarded as a black saturn \cite{saturn} in
Taub-NUT, a non-supersymmetric analogue of the solution in \cite{EEMR1}.
Upon Kaluza-Klein reduction it would describe an electric and a magnetic
black hole, both of them generically rotating under the effect of the
Poynting-induced angular momentum, and in equilibrium at a finite
distance from each other. In fact, in principle it should be possible to
introduce an \textit{arbitrary number} of black rings, which from the four-dimensional viewpoint would yield a set of electric black holes plus a magnetic one, generically non-extremal, rotating and aligned along a common axis, and in dynamical (although not thermodynamical) equilibrium.

This would be, to our
knowledge, the first known way of achieving equilibrium in an
asymptotically flat, non-supersymmetric and non-extremal multi-black
hole configuration in four dimensions. Such configurations should also be stable bound states of black holes.
We leave this interesting problem for future work.

\section*{Acknowledgements} We thank Iosif Bena, Veronika Hubeny, James
Lucietti, Mukund Rangamani, Simon Ross, and Hossein Yavartanoo for
useful discussions. PF and JC thank CEA for hospitality. JC also thanks the 
Durham University and the Hebrew University of Jerusalem for
hospitality. The work of JC and RE was supported in part by DURSI 2005
SGR 00082, MEC FPA-2007-66665-C02, and the European Community FP6
program MRTN-CT-2004-005104. JC is also supported by a MICINN FPU grant
ref AP2005-3120. A.S. was supported in part by the Marie Curie IRG 046430. PF is supported by STFC.

\appendix

\section{Supergravity solution for D6 brane with $B$-flux}
\label{app:d6b}

We want to construct a supersymmetric solution describing a D6 brane
with $B$ flux in its worldvolume. We shall consider the fluxes to be
homogeneous and isotropic in the D6 worldvolume, which for simplicity we
may consider to be wrapping a square $T^6$. In this case the solution we seek,
when lifted to M-theory and then reduced along the $T^6$, is a
solution to the minimal supergravity theory in five dimensions.

The required class of solutions has been discussed in
\cite{BD1,GMR1,EEMR1,BKW1,GSY1} (we follow mostly \cite{EEMR1}).
The solutions have metric
\beq\label{genmet}
ds^2=-Z^{-2}(dt+\omega)^2+Z h_{mn}dx^m dx^n
\eeq
with $h_{mn}$ the metric of a hyper-K\"ahler base space, and gauge potential
\beq
A=\frac{\sqrt{3}}{2}\left[Z^{-1}(dt+\omega)-\beta\right]\,.
\eeq
We take the base space to be a single-center Gibbons-Hawking space, 
\beq
h_{mn}dx^m dx^n=H^{-1}(d\psi+(\cos\theta-1)d\phi)^2+H(dr^2+r^2d\theta^2+
r^2\sin^2\theta d\phi^2)
\eeq
with
\beq\label{Hharm}
H=h+\frac{1}{r}\,.
\eeq
The solution is fully specified in terms of three more
harmonic functions $H_p$, $H_q$, $H_0$ in three-dimensional space, as
\footnote{To compare to the notation in \cite{EEMR1}, change 
$(H,H_p,H_q,H_0)\to (H_k,K,L,-2M)$, and $Z\to H$.}
\beq\label{Zfun}
Z=H_q+\frac{H_p^2}{H}
\eeq
and
\beq\label{ombe}
\omega=\omega_0(d\psi+(\cos\theta-1)d\phi)+\tilde\omega\,,\qquad
\beta=\beta_0(d\psi+(\cos\theta-1)d\phi)+\tilde\beta\,.
\eeq
with
\beq\label{ombe2}
\omega_0=-\frac{H_0}{2}+\frac{3H_p H_q}{2H}+\frac{H_p^3}{H^2}\,,\qquad 
\beta_0=\frac{H_p}{H}\,.
\eeq
The equations that determine the one-forms $\tilde\omega$ and $\tilde\beta$ in terms
of $H,H_p,H_q,H_0$ can be found in the references mentioned above. 

The residues of poles in
$(H,H_p,H_q,H_0)$ are respectively associated to numbers of D6, D4, D2,
D0 branes, so our choice for $H$ in eq.~\eqref{Hharm} corresponds to having a
single D6 brane at the origin. Indeed, when $H_p=0,H_q=1,H_0=0$ we
recover the solution for a single KK monopole. We now want to introduce
$B$-field moduli corresponding to D4 branes `dissolved' in the
worldvolume of the D6. These will also induce D0 and D2 charges, but we
do not want to introduce `pure' D0's and D2's. So we set $H_p,H_q,H_0$ to be
constant moduli $h_i$, without any poles,
\beq
H_p=h_p\,,\qquad H_q=h_q\,,\qquad H_0=h_0\,.
\eeq
With these values we easily find that $\tilde\beta=0$, and that
$\tilde\omega=\frac{h_0}{2} (\cos\theta-1)d\phi$. This last term
introduces pathological Dirac-Misner strings involving the time direction
and so it must be set to zero. Then
\beq
h_0=0\quad \mathrm{and}\quad \tilde\omega=0\,.
\eeq

Let us now rewrite
the solution in a manner convenient for KK reduction,
\beqa\label{forKK}
ds^2&=&\frac{\Sigma^2}{H^2Z^2}\left(d\psi+(\cos\theta-1)d\phi-
\frac{\omega_0 H^2}{\Sigma^2}dt\right)^2\nonumber\\
&&+\frac{HZ}{\Sigma}\left(-\frac{1}{\Sigma} dt^2+\Sigma(dr^2+r^2d\theta^2+
r^2\sin^2\theta d\phi^2)\right)\,.
\eeqa
where $\Sigma=\sqrt{Z^3H-\omega_0^2H^2}$ is the `entropy function'.
Since $r^2\Sigma$ vanishes at the D6 core at $r=0$, there is
no entropy associated to this configuration, as expected. The full
five-dimensional solution is in fact smooth there.

We shall restrict the remaining moduli
$h,h_p,h_q$ by demanding that the solution asymptotes to the
Kaluza-Klein monopole vacuum with asymptotic circle radius $L$. We
demand that as $r\to\infty$
\beq\label{fixmodul}
\Sigma\to \frac{L}{2}\,,\qquad H Z\to 1\,.
\eeq
This imposes two relations among the moduli, namely,
\beq
\sqrt{3h_p^2h_q^2+4 h h_q^3}=L\,,\qquad h_p^2+h h_q=1\,,
\eeq
which we solve in parametric form as in eq.~\eqref{hhphq}.

\section{Exact supersymmetric D0-D6 bound states}
\label{app:exactsusyd0d6}

It is not difficult to explicitly construct exact supergravity solutions
for supersymmetric bound states of D0 and D6 branes. This was first done
in an explicit manner in \cite{BD1}. Such solutions
only provide equilibrium configurations, with the distance between the
branes fully fixed by the field $B\geq B_c$, so we cannot study how the
interaction potential changes as $B$ changes. Moreover, they do not
provide any information about configurations in which supersymmetry is
broken. Nevertheless, for completenes, and as a check on our approximate
methods, we present here the
configuration with one D6 brane and $n_0$ D0 branes. It takes the form of
eqs.~\eqref{genmet}-\eqref{ombe2}
but now
\beqa
H&=&\frac{2}{L}\frac{1-3b^2}{\sqrt{(1+b^2)^3+2b
n_0(3-b^2)+n_0^2}}+\frac{1}{r}\,,\nonumber\\
H_p&=&\frac{2b(1+b^2)+n_0}{\sqrt{(1+b^2)^3+2b
n_0(3-b^2)+n_0^2}}\,,\nonumber\\
H_q&=&\frac{L}{2}\frac{(1+b^2)^2+2bn_0}{\sqrt{(1+b^2)^3+2b
n_0(3-b^2)+n_0^2}}\,,\nonumber\\
H_0&=&\frac{L^3}{8}\left(\frac{2}{L}\frac{(1-3b^2)n_0}{\sqrt{(1+b^2)^3+2b
n_0(3-b^2)+n_0^2}}+\frac{n_0}{\sqrt{r^2+R^2-2rR\cos\theta}}\right)
\eeqa
and the distance between the D0 and D6 is fixed to $r=R$, with
\beq\label{exactR}
R=\frac{L}{2}\frac{\sqrt{(1+b^2)^3+2b n_0(3-b^2)+n_0^2}}{3b^2-1}\,.
\eeq
Observe that when $n_0=0$ we reproduce the same background as in the
previous subsection. Moreover, the equilibrium distance \eqref{equilR},
obtained in the limit where the backreaction from the D0 branes, 
agrees with the exact result \eqref{exactR} when $n_0\to 0$.

\section{Maison data}
\label{sec:appMaison}
In this appendix we present the computation of the Maison data $(\chi,\kappa)$ of our seed solution \eqref{eqn:metricxy}. To avoid cluttering formulas, we will set $\ell=1$ and we will restore the units when needed.

As a first step, we compute the quantities $\lambda_{ab}$, $\tau$ and $\omega^a=\omega^{a}_{\phantom 0-}d\phi_-$, needed to rewrite the seed metric (\ref{eqn:metricxy}) in the form (\ref{seedmaison}). This only requires algebraic manipulations, and one finds:
\begin{subequations}
\begin{align}
\lambda_{00}&=-\frac{H(y,x)}{H(x,y)}\,,\\
\lambda_{01}&=-\frac{R\,C_1}{2H(x,y)}\,\big(\omega_\psi+\bar b_4\,C_2\,\omega_\phi\big)\,,\\
\lambda_{11}&=\frac{R^2}{4(x-y)^2}\,\frac{1}{H(y,x)}\left[F(x,y)-F(y,x)+2J(x,y)
	-\frac{C_1^2(x-y)^2}{H(x,y)}\big(\omega_\psi+\bar b_4\,C_2\,\omega_\phi\big)^2\right]\,,\\
 \tau&=\frac{R^2}{4(x-y)^2}\,\frac{1}{H(y,x)}\big[F(x,y)-F(y,x)+2J(x,y)\big]\,,\label{tauseed}\,\\
  \omega^{0}_{\phantom 0-}&=\frac{R\,C_1}{H(y,x)}\,
	\frac{\omega_\psi\big[F(x,y)+J(x,y)\big]
	+\bar b_4\,C_2\,\omega_\phi\big[F(y,x)-J(x,y)\big]}{F(x,y)-F(y,x)+2J(x,y)}\,,\\
  \omega^{1}_{\phantom 1-}&=\frac{F(x,y)+F(y,x)}{F(x,y)-F(y,x)+2J(x,y)}\,.
 \end{align}
\end{subequations}

To compute the scalar potentials $V_a$ and the matrix $\kappa$, one needs instead to solve differential equations, which is, in practice, a difficult task in our case. However it was shown in \cite{GS2} that one can relate $V_a$ and $\kappa$ to some auxiliary matrices, $\Gamma$ and $\tilde\kappa$, which can be computed via a generalization of BZ techniques.

\subsubsection*{The $\Gamma$ and $\tilde\kappa$ matrices}
The $\Gamma_0$ and $\tilde\kappa_0$ matrices of the solution $\tilde G_0$ are given by
\begin{eqnarray}
 \Gamma_0&=&\frac{1}{2}\,\textrm{diag}\left\{\mu_3-\bar\mu_1-\bar\mu_4\,,\bar\mu_2-\bar\mu_1\,,
	\bar\mu_3-\bar\mu_2\right\}\,,\\
\tilde\kappa_0&=&\frac{1}{8}\,\textrm{diag}\left\{\mu_3^2-\bar\mu_1^2-\bar\mu_4^2\,,\bar\mu_2^2-\bar\mu_1^2\,,
	\bar\mu_3^2-\bar\mu_2^2\right\}\,.
\end{eqnarray}
To construct the $\Gamma$ and $\tilde\kappa$ matrices of the seed solution \eqref{eqn:finalG} one can use the fact that the metric $G$ can be constructed in a two-step process,
\begin{equation}
 G=\mu_4\left(\mathbf 1-\frac{\rho^2+\mu_4^2}{\mu_4^2}\,\mathbf P_2\right)
	\left(\mathbf 1-\frac{\rho^2+\bar\mu_1^2}{\bar\mu_1^2}\,\mathbf P_1\right)\tilde G_0\,
\end{equation}
where $\mathbf 1$ is the $3\times 3$ identity matrix, and $\mathbf P_{1,2}$ are the projectors 
\begin{equation}
(\mathbf P_i)_{ab}=\frac{m_c^{(i)}(G_0)_{ca}\,m_b^{(i)}}{m_d^{(i)}(G_0)_{df}\,m_f^{(i)}}\;.
\end{equation}
In the equation above $G_0$ denotes the seed metric at each step and $m_a^{(i)}$ are the vectors constructed out of the BZ vectors $m_{0a}^{(i)}$ and the seed solution at each step,
\begin{equation}
m_a^{(i)}=m_{0b}^{(i)}\left[\Psi^{-1}_0(\mu_i,\rho,z)\right]_{ba}\;.
\end{equation}

Defining a new matrix $\mathbf Q_i$ as
\begin{equation}
 \mathbf Q_i=\frac{\rho^2+\mu_i^2}{\mu_i}\,\mathbf P_i\,,
\end{equation}
one can show that the $\Gamma$ and $\tilde\kappa$ matrices of the solution $G$ are given by
\begin{eqnarray}
 \Gamma&=&\Gamma_0+\frac{1}{2}\sum_{i=1}^2\mathbf Q_i+\frac{1}{2}\,\bar\mu_4\mathbf 1\,,\\
\tilde\kappa&=&\tilde\kappa_0+\frac{1}{8}\left(\frac{\bar\mu_1^2-\rho^2}{\bar\mu_1}\,\mathbf Q_1+\frac{\mu_4^2-\rho^2}{\mu_4}\,\mathbf Q_2\right)
+\frac{1}{4}\sum_{i=1}^2\big[\mathbf Q_i,\Gamma_0\big]
-\frac{1}{8}\big[\mathbf Q_1,\mathbf Q_2\big]+\frac{1}{8}\,\bar\mu_4^2\,\mathbf 1\;.
\end{eqnarray}
The explicit expressions of the $\Gamma$ and $\tilde\kappa$ matrices are too involved to be written down here. 

\subsubsection*{$V_a$ and $\kappa$}
The potentials $V_a$ can be computed from the matrix $\Gamma$ derived above as
\begin{equation}
 V_0=\Gamma_{0}^{\phantom 0-}+c_0\,,~~~~~~ V_1=\Gamma_{1}^{\phantom 1-}+c_1\,,
\end{equation}
where the subscripts $\pm$ denote components in the base \eqref{eqn:newcoords}, and $c_{0,1}$ are constants which are determined by the asmptotic boundary conditions. As discussed in \cite{GS1,GS2}, the matrix $\chi$, computed from these data,  should approach a constant matrix $\eta_5$ at asymptotic spatial infinty,  
\begin{equation}
 \chi \xrightarrow{r\to \infty} \eta_5\equiv\left(
\begin{array}{ccc}
 -1 & 0 & 0\\
0 & 0 & 1\\
0 & 1 & 0
\end{array}
\right)\,. 
\label{eqn:eta5}
\end{equation}
In the $(x,y)$ coordinates that we are using, spatial infinity lies at $x\to y \to -1$, and the correct asymptotics \eqref{eqn:eta5} is achieved by taking
\begin{equation}
 c_0=0\,,~~~~~~c_1=\frac{R^2(\lambda-2\nu+\lambda \nu)}{4(1-\lambda)}\,.
\end{equation}
One can also check that the sub-leading correction to the asymptotic limit of the $\chi$ matrix is of the form 
\begin{equation}
\chi=\eta_5\left[\mathbf 1-\frac{\delta\chi}{r^2}+O\left(\textstyle{\frac{1}{r^4}}\right)\right]\,,
\end{equation}
where $\delta\chi$ is a constant $3\times 3$ matrix. The matrix $\delta\chi$ contains the information about the conserved charges of the solution.

The last piece of data needed is the matrix of one-forms $\kappa$. The components of this matrix can be derived from the following relations \cite{GS2}:
\begin{equation}
\begin{aligned}
\kappa_{00}&=V_0\,\omega^0+\Gamma_{0}^{\phantom 00}+c_{00}\,,\\
\kappa_{01}&=V_1\,\omega^0+\Gamma_+^{\phantom +0}+c_{01}\,,\\
\kappa_{02}&=-\omega^0+c_{02}\,,\\
\kappa_{10}&=V_0\,\omega^1+\Gamma_{0}^{\phantom 0+}+c_{10}\,,\\
\kappa_{11}&=V_1\,\omega^1+\Gamma_{+}^{\phantom++}+c_{11}\,,\\
\kappa_{12}&=-\omega^1+c_{12}\,,\\
\kappa_{20}&=V_0\left(V_0\,\omega^0+V_1\,\omega^1\right)+\frac{1}{2}\left(\Gamma\sigma\Gamma\right)_0^{\phantom 0-}+\tilde\kappa_0^{\phantom 0-}+c_0\left(\Gamma_{0}^{\phantom 00}-\Gamma_{-}^{\phantom--}-z\right)+c_1\,\Gamma_{0}^{\phantom 0+}+c_{20}\,,\\
\kappa_{21}&=V_1\left(V_0\,\omega^0+V_1\,\omega^1\right)+\frac{1}{2}\left(\Gamma\sigma\Gamma\right)_+^{\phantom +-}+\tilde\kappa_+^{\phantom +-}+c_1\left(\Gamma_{+}^{\phantom ++}-\Gamma_{-}^{\phantom--}-z\right)+c_0\,\Gamma_{+}^{\phantom +0}+c_{21}\,,\\
\kappa_{22}&=-V_0\,\omega^0-V_1\,\omega^1-\Gamma_{0}^{\phantom 00}-\Gamma_{+}^{\phantom ++}-c_{00}-c_{11}\,,
\end{aligned}
\end{equation}
where $\sigma$ is a constant matrix which, in the $(t,\phi_+,\phi_-)$ basis, is given by
\begin{equation}
\sigma=\left(
\begin{array}{ccc}
1 & 0 & 0\\
0 & 1 & 0\\
0& 0 & -1
\end{array}\right)\,.
\end{equation}
The $c_{ij}$'s are constants that are fixed by requiring that
asymptotically, the $\kappa$ matrix approaches
\begin{equation}
\kappa\approx -\frac{\delta\chi}{4}\,\cos2\theta\,d\phi_-+O\left(\textstyle{\frac{1}{r^2}}\right)\,.
\end{equation}
For our solution \eqref{eqn:metricxy}, we can obtain the correct
asymptotics for the $\kappa$ matrix by fixing these constants as
\begin{equation}
\begin{aligned} 
c_{00}&=\,\frac{R^2\lambda(1-\nu)}{2(1-\lambda)}\,,\\
c_{01}&=\,c_{20}=-\frac{R^3C_1(1-\nu ) \big[(1-\lambda +\bar b_4 C_2  (1-\nu ) (1+\nu-2\lambda\nu )\big]}{4 (1-\lambda ) \left[1-\lambda +\bar b_4 C_2 (1+\lambda ) (1-\nu )^2\right]}\,,\\
c_{02}&=\,c_{10}=\frac{R\,\bar b_4 C_1C_2  (1-\nu )^2}{1-\lambda +\bar b_4 C_2   (1+\lambda ) (1-\nu )^2}\,,\\
c_{11}&=-\frac{R^2\big[\,\bar b_4(\lambda-\nu)(1-\nu)^2+C_2\,\nu(1-\lambda)\big]}
	{2(1-\lambda)\big[\bar b_4(1-\nu)^2+C_2\big]}\,,\\
c_{12}&=\,0\,,\\
c_{21}&=-\frac{R^4\lambda(\lambda-\nu)(1-\nu)}{8(1-\lambda)^2}\,.
\end{aligned}
\end{equation}
Once this constants are fixed, the $\kappa$ matrix is uniquely
determined; the explicit expression is very long and we will not give
it.

\section{Conserved charges}
\label{app:charges}

Upon reduction along the KK circle parametrized by the coordinate $\xi^1$ in \eqref{eqn:finalsol} we obtain a four dimensional asymptotically flat solution. This solution consists in an KK electrically charged rotating black hole separated from the nut, which accounts for the magnetic charge.  We can easily compute the conserved charges of the four-dimensional solution, and we find:
\begin{subequations}
\begin{align}
M_\mathrm{tot}&=\frac{\ell}{16\,G_4\,(1+b_4)(1-\nu^2)^2}\,\bigg\{
(1+\nu)^2\Big[D_1\,c_{2\gamma}(3+\ch_{2\alpha})-6\big(D_1-4\,\hat R^2\,\nu(1-\nu)(1+b_4)\big)\,\sh^2_\alpha\nonumber\\
&\hspace{6.7cm}+4\,D_3\,s_\gamma\,\sh_{2\alpha}\Big]\nonumber\\
&\hspace{3.9cm}
+\hat R\,D_\nu\,\nu(1-\nu)\Big[\,D_2\,s_{2\gamma}(3+\ch_{2\alpha})
	+2\big(D_2+8\,b_4(1-\nu)\big)\,c_\gamma\,\sh_{2\alpha}\Big]
\bigg\}\,,\\
Q&=\frac{\ell}{8(1+b_4)(1-\nu^2)^2}\,\bigg\{\Big[(1+\nu)^2D_1\,c_{2\gamma}-3(1+\nu)^2\big(D_1-4(1+b_4)\hat R^2\,\nu(1-\nu)\big)\nonumber\\
&\hspace{4.5cm}+\,\hat R\,D_\nu\,D_2\,\nu(1-\nu)\,s_{2\gamma}\Big]\sh_{2\alpha}\nonumber\\
&\hspace{3.9cm}+\Big[4\,
(1+\nu)^2\,D_3\,s_\gamma+2\,\hat R\, D_\nu\,\nu(1-\nu)\big(D_2+8\,b_4(1-\nu)\big)c_\gamma
\Big]\,\ch_{2\alpha}
\bigg\}\,,\\
P&=\frac{\ell}{8 | 1+b_4 |(1-\nu^2)^2}\,\bigg\{
2\Big[\hat R\,D_\nu\,D_2\,\nu(1-\nu)\,c_{2\gamma}-(1+\nu)^2D_1\,s_{2\gamma}\Big]\,\sh_\alpha\nonumber\\
&\hspace{3.9cm}+\Big[\!- 2\,\hat R\,D_\nu\,\nu(1-\nu)\big(D_2+8\,b_4(1-\nu)\big)\,s_\gamma+ 4\,(1+\nu)^2D_3\,c_\gamma\Big]\,\ch_\alpha
\bigg\}\,,\\
J&={R^2\over 16\, G_4\,(1+b_4)^2 (1-\nu)^3 (1+\nu)^2}\,\bigg\{\Big[D_7\, (1+\nu)\, s_{2\gamma}+2\,\hat R\,D_\nu\,D_8\,(1-\nu)\,\nu\,c_{2\gamma}\Big] \,\ch_\alpha\nonumber\\
&\hspace{3.7cm}-2\,\Big[\hat R\,D_\nu\,(D_8 + D_9) (1-\nu)\,\nu \, s_\gamma + (D_7 +D_{10}\,\hat R^2 \,\nu^2)(1+\nu)\,c_\gamma\Big]\sh_\alpha\bigg\}\,,
\end{align}
\label{eqn:conservedCH}
\end{subequations}
Similarly, the horizon area, temperature and angular velocity are found to be
\begin{subequations}
\begin{align}
 \mathcal{A}_4&={\pi\, R^2 \,\nu^2\over 2 (1+b_4)^2 (1-\nu^2)^{5/2}}\,\bigg\{\Big[ 2\, \hat{R}\, D_\nu\, D_4\,(1-\nu) (1+\nu + b_4 (1-\nu))-D_5\, (1+b_4) (1+\nu)^2 s_{2\gamma} -\nonumber\\
&\hspace{4.8cm}+2\, \hat{R}\, D_\nu (1+b_4)(1-\nu^2) (D_4-2\,\nu\, \hat{R}^2 (1+\nu+b_4(1-\nu))) c_{2\gamma} \Big]\ch_\alpha\nonumber\\
&\hspace{4.6cm} +\Big[(1+\nu)^2 D_6\, c_\gamma + 4\, \hat{R} \,D_\nu\, \nu (1-\nu) (D_2+8\, b_4(1-\nu)) s_\gamma \Big]\sh_\alpha\bigg\}\;,\\
T_\textrm{H}&= {2 (1+b_4)^2 (1-\nu^2)^{5/2}\over \ell\,\pi\,\nu}\,\bigg\{\Big[ 2\, \hat{R}\, D_\nu\, D_4\,(1-\nu) (1+\nu + b_4 (1-\nu))-D_5\, (1+b_4) (1+\nu)^2 s_{2\gamma} -\nonumber\\
&\hspace{4.8cm}+2\, \hat{R}\, D_\nu (1+b_4)(1-\nu^2) (D_4-2\,\nu\, \hat{R}^2 (1+\nu+b_4(1-\nu))) c_{2\gamma} \Big]\ch_\alpha\nonumber\\
&\hspace{4.6cm} +\Big[(1+\nu)^2 D_6\, c_\gamma + 4\, \hat{R} \,D_\nu\, \nu (1-\nu) (D_2+8\, b_4(1-\nu)) s_\gamma \Big]\sh_\alpha\bigg\}^{-1}\;,\label{eqn:thetH}\\
\Omega_\textrm{H}&={8 |1+b_4| (1-\nu^2)^2\over\ell} \,\bigg\{\Big[ 2\, \hat{R}\, D_\nu\, D_4\,(1-\nu) (1+\nu + b_4 (1-\nu))-D_5\, (1+b_4) (1+\nu)^2 s_{2\gamma} -\nonumber\\
&\hspace{4.8cm}+2\, \hat{R}\, D_\nu (1+b_4)(1-\nu^2) (D_4-2\,\nu\, \hat{R}^2 (1+\nu+b_4(1-\nu))) c_{2\gamma} \Big]\ch_\alpha\nonumber\\
&\hspace{4.6cm} +\Big[(1+\nu)^2 D_6\, c_\gamma + 4\, \hat{R} \,D_\nu\, \nu (1-\nu) (D_2+8\, b_4(1-\nu)) s_\gamma \Big]\sh_\alpha\bigg\}^{-1}\;,\label{eqn:omegaH}
\end{align}
\label{eqn:horizonCH}
\end{subequations}

To simplify the expressions for the various magnitudes in equations
\eqref{eqn:conservedCH}-\eqref{eqn:horizonCH}, we have defined
$c_\gamma\equiv \cos\gamma$, $s_\gamma\equiv \sin\gamma$, and
$\ch_\alpha\equiv \cosh\alpha$, $\sh_\alpha=\sinh\alpha$, which in turn
are fixed according to \eqref{gammasolbis} and \eqref{alphasol}
respectively. Similarly, the constants $D_i$ are given by
\begin{subequations}
\begin{align}
D_\nu&=(1+\nu)\sqrt{\frac{1+\nu}{1-\nu}}\;,\\
  D_1&=(1-b_4)- \big[2 (1-b_4)-3 (1+b_4)\, \hat R^2\big]\, \nu 
+\big[ (1-b_4)-3\, (1+b_4) \hat R^2-\textstyle{\frac{1}{2}}\displaystyle(1+b_4)\hat R^4\big]\, \nu ^2\nonumber\\
&~~~~-\textstyle{\frac{1}{4}}(1-b_4)\,\hat R^4\, \nu ^4\;,\\
D_2&=\hat R^2(1+\nu)^2-4\,b_4\big[1-\nu-\textstyle{\frac{1}{4}}\hat R^2(1+2\,\nu-\nu^2)\big]\;,\\
D_3&=(1-b_4)\big[(1-\nu)^2+\textstyle{\frac{1}{4}\hat R^4\,\nu^4}\big]+\textstyle{\frac{1}{2}}(1+b_4)\hat R^4\,\nu^2\;.\\
D_4&=4\,(1-\nu)+\hat R^2\,(1+b_4 +(1-b_4)\,\nu^2)\\
D_5&=4\Big[ (1-\nu)^2 -3\, \hat R^2 (1-\nu)\,\nu +{1\over 4} \hat R^4\, \nu^4 \Big]-{1\over 2} (1+b_4)\, \hat R^4\, (1-\nu^2)^2\\
D_6&=(1+b_4)\Big\{8 \Big[(1-\nu)^2 + \hat R^2 \,(1-\nu) -{1\over 4} \hat R^4 \,\nu^4 \Big] +(1+b_4) \,\hat R^4 \,(1-\nu^2)^2  \Big\} \nonumber\\
&\hspace{.5cm} + 8\, (1-b_4)\, \hat R^2\, (1-\nu)\,\nu^2\\
D_7&={1\over 2}(1+\nu)\, [4\,(1-\nu)^2 (1-\nu+2\nu^2) -12\, \hat R^2 \,(1-\nu^2)\,\nu^2 -\hat R^4\, (1-2\nu-\nu^3)\,\nu^3]\nonumber\\
&\hspace{.5cm}-b_4 \hat R^4 (1-\nu^2)^2\nu^3 
-{1\over 2}b_4^2\, (1-\nu)\,[4\,(1-\nu)^2 (1+\nu+2\nu^2) -12\,\hat R^2 \,(1+2\nu-3\nu^2)\,\nu^2 \nonumber\\
&\hspace{1cm}+\hat R^4\, (1+2\nu+\nu^3)\,\nu^3]\\
D_8&=2\,\hat R^2\,(1+\nu^2)\,\nu^2-2\,(1-b_4^2)\,(1+\nu)^2\,(1-\nu)-2\, b_4^2\,(1-\nu)\,\nu^2\,(4+\hat R^2(1+\nu))\nonumber\\
&\hspace{.5cm} -(1+b_4)^2\,\hat R^2\, (1-\nu)^2\,\nu^2\\
D_9&= 4\,(1-\nu)\, [(1+\nu)^2\,(1-b_4^2) +4\, b_4^2\, \nu^2]\\
D_{10}&=(1+\nu)\,[6\,(1-\nu^2) +\hat R^2 \,(1-2\nu-\nu^3)\,\nu] + 2\, b_4 \,\hat R^2\, (1-\nu^2)^2\,\nu\nonumber\\
&\hspace{.5cm}-b_4^2\, (1-\nu)\,[6\,(1+2\nu-3\nu^2)-\hat R^2\, (1+2\nu+\nu^3)\,\nu]
\end{align}
\label{eqn:constants}
\end{subequations}

We have checked numerically that if the parameters $\nu$ and $\hat R$
are constraint to vary in the ranges \eqref{eqn:ranges}, the mass $M$,
the temperature $T_\textrm{H}$ and the horizon area ${\cal
A}_\textrm{H}$ are always positive.

In terms of the four dimensional quantities, the five-dimensional
angular momenta corresponding to the original angles
\eqref{eqn:newcoords} are given by
\begin{equation}
J_{\hat{\psi}}^{(5)}=\frac{P\,Q}{G_4}+J\;,\qquad J_{\hat{\phi}}^{(5)}=\frac{P\,Q}{G_4}-J\;.
\label{eqn:4d5d}
\end{equation}
Note that the angles $\hat{\psi}$ and $\hat{\phi}$ are related to the angles $\psi$ and $\phi$ used
in sec. \ref{sec:pertd0d6} as
\be
\hat{\psi}={\psi\over 2}\,,\quad \hat{\phi}={\psi\over 2}-\phi\,,
\ee
and thus the corresponding conserved charges are related as
\be
J_\psi= {1\over 2}(J_{\hat{\psi}}^{(5)}+J_{\hat{\phi}}^{(5)})=\frac{P\,Q}{G_4}\,,\quad J_\phi = -  J_{\hat{\phi}}^{(5)}=J-\frac{P\,Q}{G_4}\,.
\label{jrelation}
\ee

The horizon area of the five-dimensional solution is given by
\begin{equation}
\mathcal A_5=2\pi\,L\,\mathcal{A}_4\;, \label{eqn:aH5d}
\end{equation}
where $L$ is defined below \eqref{eqn:theell}. The temperature of the
horizon of the four- and the five-dimensional solutions coincides. The
mass of the five-dimensional solution can be computed as
\begin{equation}
M=M_\mathrm{tot}-M_\mathrm{D6}\;,
\end{equation}
where $M_\mathrm{D6}$ is the mass of the KK monopole \eqref{eqn:KKmass}.

Finally we notice that regularity of the \textit{new} five-dimensional
metric \eqref{eqn:finalsol} imposes that $\xi^1$ has to be periodically
identified as
\begin{equation}
\xi^1\sim \xi^1+2\pi L\;,\qquad L=\frac{4P}{N_6}\;, \label{eqn:theell}
\end{equation}
for an integer $N_6$.

\section{Physical magnitudes of the approximate doubly spinning solution}
\label{app:doublespin}

It is very easy to include in our perturbative construction a second
independent angular momentum along the $S^2$ of the ring. We need only
use the stress tensor that reproduces the long-distance field of a
boosted Kerr black string. This turns out to be very simple, since it
takes the same form as \eqref{tensorcomponents}, plus an additional
component $T_{\tau\phi}$ for the spin along $\phi$. Actually, we do not
even need the details of this component, since it falls off sufficiently
fast at infinity so as to not affect the equilibrium equations. To
obtain the value of the spin $J_\phi$ of the ring, we need simply
compute it for the boosted Kerr string at the equilibrium boost.

Then, the five-dimensional physical magnitudes of the approximate doubly
spinning black ring in Taub-NUT (so $b=0$) at equilibrium (so
$\sinh^2\alpha=1$) are found to be
\begin{subequations}
\begin{align} 
M&=\frac{3}{4\,G_5}\,r_0\,\Delta z\;,\\
J_{\psi}&=\frac{\sqrt 2}{16\pi\,G_5}\,r_0\,(\Delta z)^2\;,\qquad
J_\phi=\frac{\sqrt 2}{2G_5}\,r_0\,a\,\Delta z\;,\label{eqn:5dmom}\\
{\mathcal A}_5&=4\pi\sqrt 2\,(r_+^2+a^2)\Delta z\;,
\end{align}
\label{perturbativecharges}
\end{subequations}
where $r_+=2r_0+\sqrt{(2r_0)^2-a^2}$ and $a$ is the Kerr rotation parameter.
The four-dimensional magnitudes are obtained using 
\beq\label{qjpsi}
Q=\frac{G_4 J_\psi}{P}\,,\qquad J=J_\psi+J_\phi\,,
\eeq
as follows from \eqref{chipsi}, \eqref{barphi}, and \eqref{Qel}. The four-dimensional area is
obtained as in \eqref{eqn:aH5d}. Clearly, $G_4 J\neq QP$ if $a\neq 0$.

For the exact solution, the configuration in which the ring is far from
the NUT is described by the limit in which $\nu\to 0$ and
$\hat{R}=\sqrt{2}(1-\eta \nu)$, for some fixed $\eta$. In this limit the
mass, charges and area of the exact solution become
\bea
{G_4 M_\text{tot}\over P} &=&{3+\eta\over 2 \eta}-{18\eta^2-16\eta+3\over 8 \eta^2}\nu\,,\nonumber\\
{Q\over P}&=& {\sqrt{2}\over\eta}-{10\eta^2-12\eta+1\over 2\sqrt{2}\eta^2}\nu\,,\nonumber\\
{G_4 J\over P^2}&=& {\sqrt{2}\over\eta}-{10\eta^2-12\eta-3\over 2\sqrt{2}\eta^2}\nu\,,\nonumber\\
{\mathcal{A}_4\over P^2}&=& {16\sqrt{2}\pi\over \eta^2}-8\sqrt{2}\pi {4\eta^2-2\eta+1\over \eta^3}\nu\,.
\eea
It can be checked that these values match the ones of the perturbative
solution given in (\ref{perturbativecharges}) after the following
reparametrization:
\be
\Delta z = 2\pi L \Bigl(1-{3\eta-5\over 3}\nu\Bigr)\,,\quad
r_0={L\over 2\eta}\Bigl(1-{6\eta^2+4\eta+3\over 12\eta}\nu\Bigr)\,,\quad
a= {\nu\over 2\eta}\,.
\label{mapexactperturbative}
\ee
One can also check that the exact metric reduces, at first order in
$\nu$ and after some change of coordinates, to that of a boosted Kerr
black string, with the angular momentum parameter $a$ given in
(\ref{mapexactperturbative}) and boost parameter given by
$\sinh\alpha=1+2\nu$.

\end{document}